\documentclass[aps,pra,floatfix,twocolumn,footinbib,amsmath,amssymb, superscriptaddress]{revtex4-1}

\usepackage                             {epsfig}
\usepackage                             {natbib}
\usepackage                             {xspace}
\usepackage                             {amsmath}
\usepackage                             {graphicx}
\usepackage     [english]               {babel}
\usepackage                             {natbib}
\usepackage                             {hyperref}
\usepackage                             {amsfonts}
\usepackage                             {amssymb}
\usepackage                             {latexsym}
\usepackage								{bibentry}
\usepackage		                {color}
\usepackage       			{ulem}
\usepackage{multibib}

\definecolor{darkgreen}{rgb}{0,.5,0}

\newcommand{\figref}[1]{Fig.~\ref{#1}}

\renewcommand{\eqref}[1]{Eq.~(\ref{#1})}
\newcommand{\ket}[1]{| #1 \rangle}

\newcommand{\braket}[1]{\langle #1 \rangle}

\newcommand{\ignore}[1]{}
\newcommand{\nobibentry}[1]{{\let\nocite\ignore\bibentry{#1}}}
\newcommand{\bibfnamefont}[1]{#1}
\newcommand{\bibnamefont}[1]{#1}
\begin{document}
\title{Inelastic Confinement-Induced Resonances in Quantum Dots} 
\author{Maria Troppenz, Simon Sala, Philipp-Immanuel Schneider, and Alejandro Saenz}

\affiliation{AG Moderne Optik, Institut f\"ur Physik,
  Humboldt-Universit\"at zu Berlin, Newtonstrasse 15,
  12489 Berlin, Germany}
\date{\today}

\begin{abstract}
  Recently, it was shown that the coupling of center-of-mass and
  relative motion in atomic systems leads to inelastic
  confinement-induced resonances (ICIRs) [\textit{Phys.~Rev.~Lett.}
  \textbf{109}, 073201 (2012)].  In the present work, the possible occurrence 
  of ICIRs in quantum dots is investigated. Particularly,
  electron-hole and electron-electron two-body systems with long-range
  Coulomb interaction are considered using the material parameters of
  GaAs. ICIRs are identified for the electron-hole system verifying
  the universal nature of the ICIR and, additionally,
  resonances due to the coupling of center-of-mass and relative 
  motion are found also for the electron-electron system. 
  In analogy to the coherent molecule
  formation appearing at ICIR in atomic systems a significant change
  in the mean distance between electrons and holes at the resonance is
  observed. By using the redistribution of the particle densities at the
  resonance position in modern quantum-dot experiments, the ICIR can
  provide a new technique for the control of the electron distribution in
  quantum dots and for the generation of single photons on demand.
\end{abstract}
\maketitle
%
%
%
%
\section{Introduction}
 Artificial atoms and quantum dots (QDs) have attracted a great deal of
 attention during recent years \cite{qd:bera10}. Their modifiable 
 physical properties offer a variety of new technologies in electronics and
 optoelectronics, such as single-photon
 sources \cite{qd:eisa11,qd:clau10} or single-electron
 transistors \cite{qd:devo00,qd:stam08}. In QDs, the charge
 carriers, electrons and holes, are strongly confined in each spatial
 direction by an external potential leading to a discretization of the
 energy spectrum. The external confinement can be realized by modern
 fabrication techniques like lithography and epitaxy 
 or electrostatically.

 While in theoretical treatments a simplified external potential such
 as a box potential \cite{qd:aliv96,qd:unlu06} or a harmonic potential
 \cite{qd:brey89,qd:kuma90,qd:que92} is often adopted, the
 exact shape of the confinement of a quantum dot is in general not
 known. Yet, the confinement is definitely finite and hence never purely 
 harmonic. Such an anharmonicity leads to a coupling of the 
 center-of-mass (c.m.)\ and relative (rel.)\ motion. It was 
 demonstrated recently for ultracold atoms
 \cite{cold:sala12,cold:sala13} 
 that the c.m.-rel.\ coupling allows for a controlled transfer of two atoms into
 a bound molecular state. This phenomenon is denoted as \textit{inelastic}
 confinement-induced resonance (ICIR) \cite{cold:olsh98b}.

 In the present work the appearance of resonances due to c.m.-rel.\ coupling 
 and thus ICIRs for a two-body system of Coulomb-interacting particles
 is investigated. The occurrence of ICIR is demonstrated, thus further
 demonstrating their universality. It is also shown that the absence of a
 well-defined last bound state as is the case for ultracold atoms
 leads to modifications reflecting the influence of the long-range
 Coulomb interaction.

 In fact, ICIRs occur both for attractive interactions (excitons) and for
 repulsive interactions (electron pairs). For electron pairs it is
 demonstrated that the com.-rel.\ coupling does not cause a significant
 change in the mean interparticle distance but nevertheless such a 
 coupling can result in
 instabilities and modifications of the charge distribution. In the
 case of excitons, a variation of the QD geometry allows, however, for a
 substantial modification of the mean distance between the electron and
 the hole. It is demonstrated that an adiabatic transition from an 
 unbound electron-hole pair to a more tightly bound state seems 
 experimentally feasible in an exciton system. Since
 a reduced mean distance changes the recombination rate
 \cite{qd:chri84}, ICIR are expected to deliver a novel kind of
 deterministic single-photon sources. Excitons in electrostatic traps
 might be especially suitable for this purpose, since they can be 
 manipulated \textit{in situ}.

 The paper is organized in the following way. In Sec.\ \ref{sec:icir},
 the mechanism of ICIR is introduced briefly for ultracold atoms. In
 Sec.\ \ref{sec:2} the full two-body energy spectrum of a confined
 Coulomb system and the computational approach are introduced. Sec.\
 \ref{sec:3} describes the model for the QD confinement.  The results
 are presented in Sec.\ \ref{sec:4}. This includes the discussion of
 the energy spectra in \ref{sec:enSp}, the analysis of the coupling
 strength in Sec.\ \ref{sec:coupl}, and the consequences of a variation 
 of the confinement in Sec.\ \ref{sec:conVar}. 
 Finally, the paper closes with a conclusion and outlook (Sec.\ \ref{sec:6}).

\section{Inelastic confinement-induced resonances}
\label{sec:icir}
 Recently, ICIR were discovered in the context of trapped ultracold
 atoms. Large losses of the trapped atoms for a given interatomic 
 interaction strength (tuned with the aid of magnetic Feshbach 
 resonances) were observed in \cite{cold:hall10b} and explained 
 by the occurrence of an ICIR in \cite{cold:sala12}. In a 
 dedicated experiment \cite{cold:sala13} the occurrence 
 of the ICIR was then more directly confirmed.  

\begin{figure}[ht!]
  \begin{center}
    \includegraphics[width=0.45\textwidth]{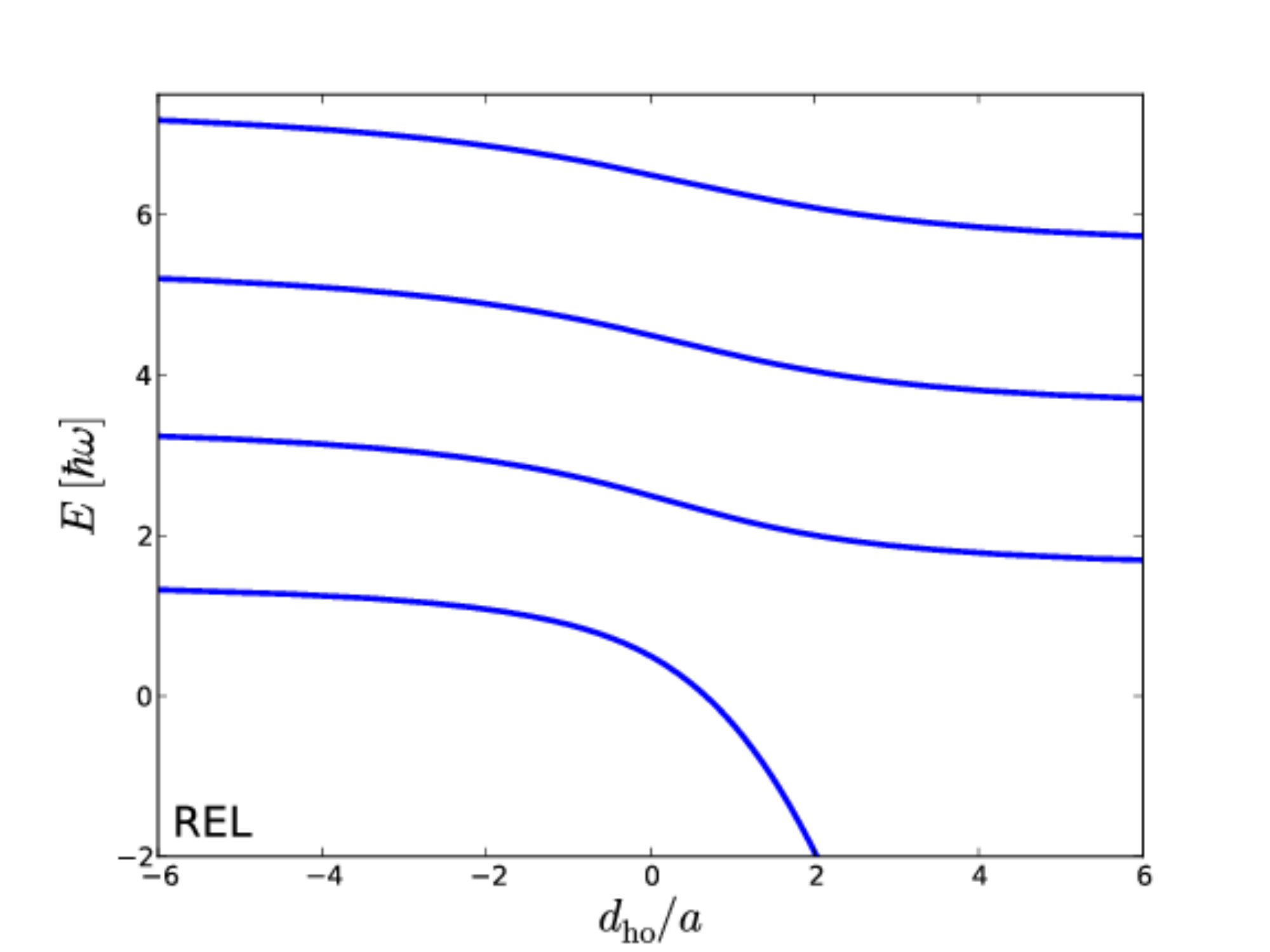}\\
    \includegraphics[width=0.45\textwidth]{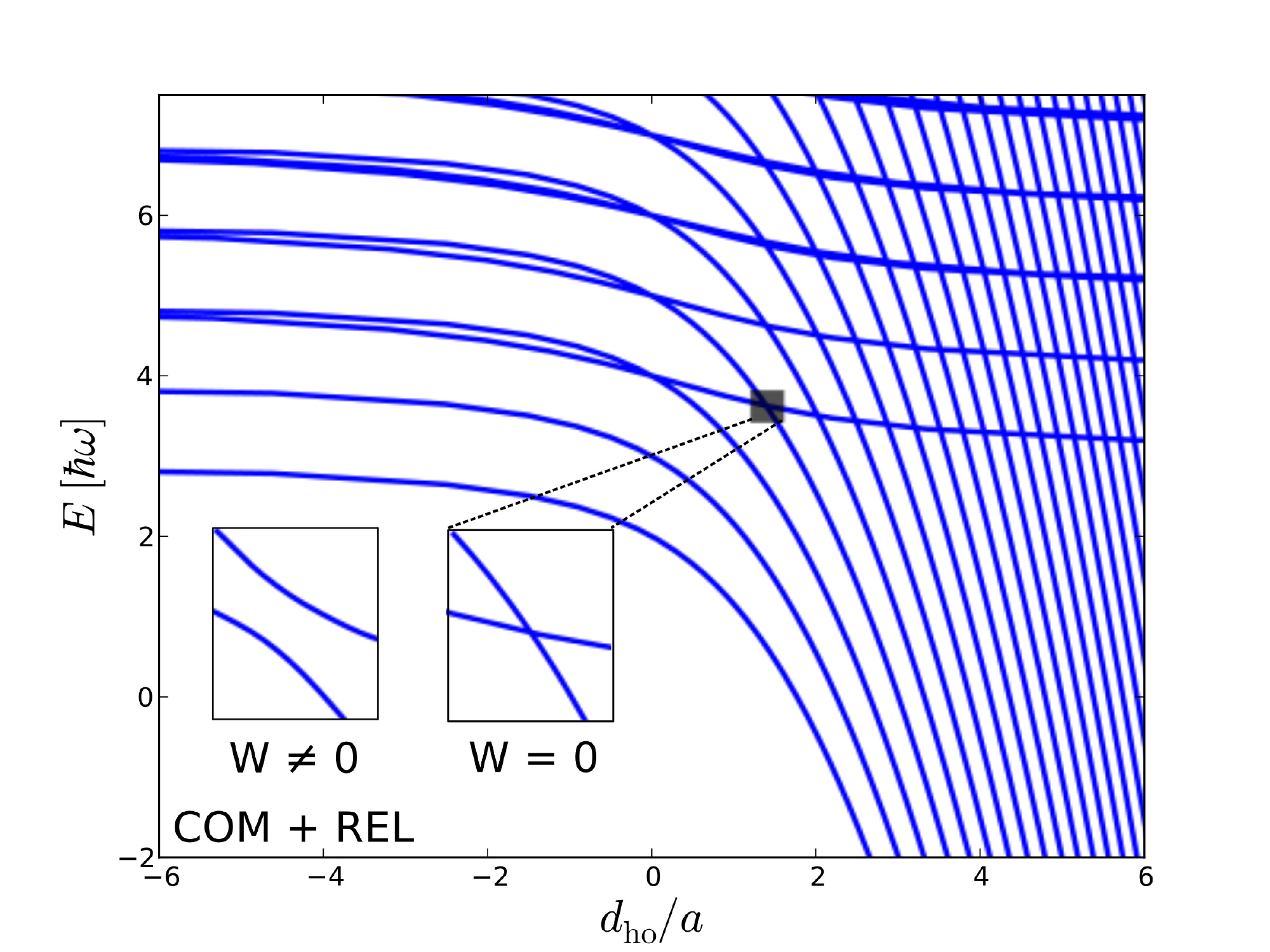}
    \caption{(Color online) Eigenenergy spectra of two atoms interacting via a
      $\delta$ pseudopotential and confined in a 3D isotropic harmonic
      potential for varying (inverse) interaction strength represented 
      by the scattering length $a$. The upper part shows the rel.\ motion 
      spectrum, while the 
      lower part shows the total spectrum including the c.m.\ motion. A
      non-vanishing coupling $W$ between c.m.\ and rel.\ motion as
      introduced by, e.\,g., an anharmonicity of the external potential
      makes the crossings between states become avoided as illustrated in the
      inset.}
    \label{fig:icir}
  \end{center}
\end{figure}

 The interatomic interaction in the ultracold regime is well
 described by the Fermi-Huang $\delta$ pseudopotential
\begin{equation}
  \label{eq:pseudopot}
  U_{\delta}(\mathbf{r})=\frac{4\pi\hbar^2a}{m} \: \delta(\mathbf{r}) \: 
                               \frac{\partial}{\partial r}r 
\end{equation}
 where $m$ is the atomic mass and $a$ the $s$-wave scattering length
 describing the collision in the limit of vanishing collision energy
 (temperature). Thus the interparticle interaction strength is fully
 characterized by the single parameter $a$. Experimentally, it is
 possible to vary this parameter to almost arbitrary values using
 magnetic Feshbach resonances \cite{cold:chin10}. The spectrum of two
 ultracold atoms confined in a harmonic trapping potential that
 interact via the $\delta$ pseudopotential is analytically solvable
 \cite{cold:busc98, cold:idzi06} and is shown in \figref{fig:icir} for 
 the example of an isotropic harmonic potential and for a variation of the
 inverse scattering length $d_{\mathrm{ho}}/a$ where $d_{\mathrm{ho}} =
 \sqrt{\hbar/(\mu \omega)}$ is the harmonic-oscillator length with the
 harmonic-oscillator frequency $\omega$ and the reduced mass $\mu = m_1
 m_2 / (m_1 + m_2)$ of the two particles with masses $m_1$ and $m_2$ 
 respectively. The reduced mass is equal to $m/2$ for the here considered
 case of atoms with equal masses.

\begin{figure}[ht!]
  \begin{center}
    \includegraphics[width=0.45\textwidth,natwidth=610,natheight=642]{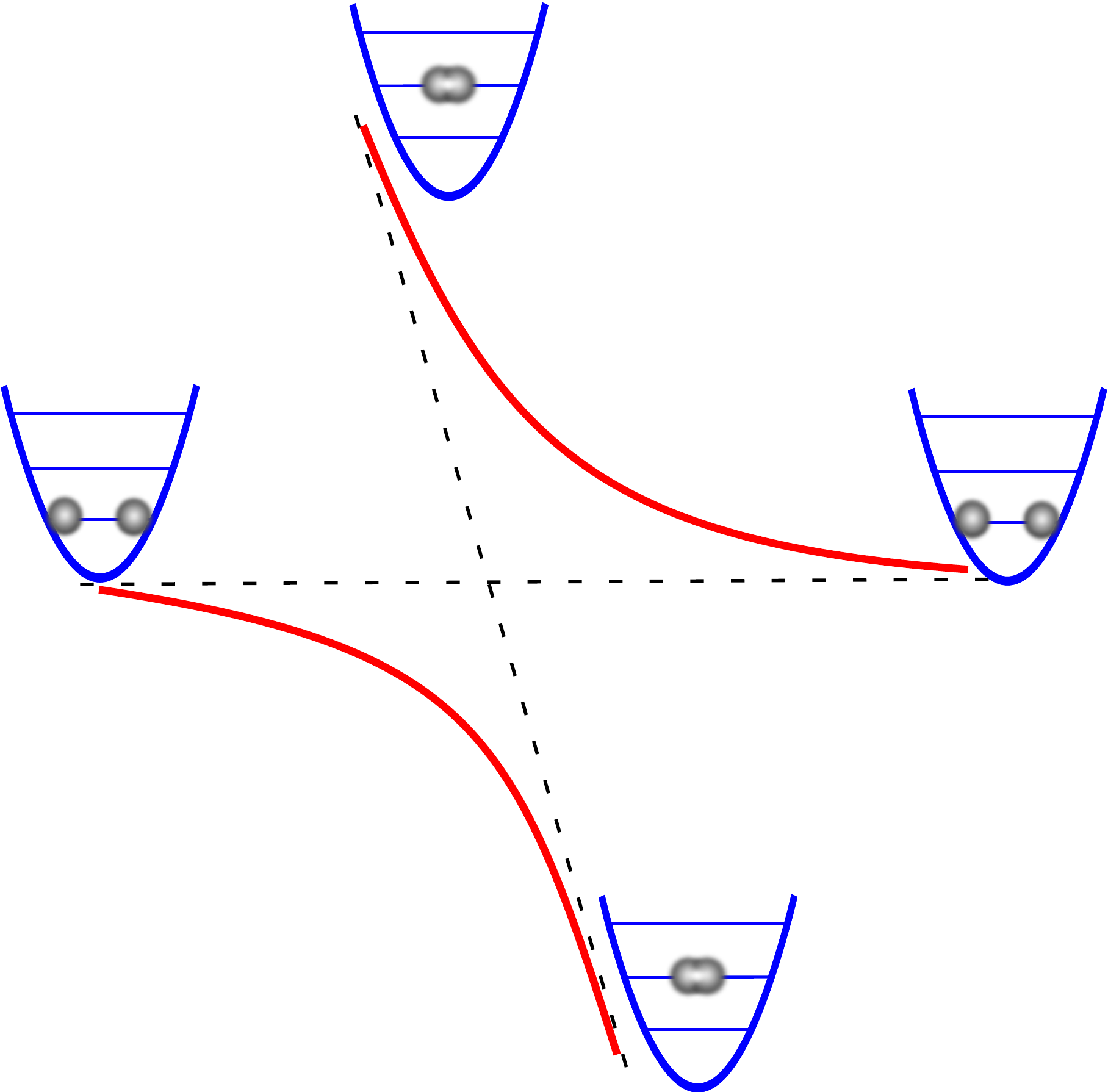}
    \caption{(Color online) Sketch of an avoided energy crossing of a molecular bound
      state with c.m.\ excitation and a trap state in the c.m.\ ground
      state. Passing through the crossing adiabatically (on the red
      solid lines) allows for an transformation of the bound state into
      a trap state and {\it vice versa}. The black dashed lines indicate 
      the diabatic curves.}
   \label{fig:sketch}
  \end{center}
\end{figure}

 A characteristic feature of the rel.\ motion spectrum is the occurrence 
 of a shallow molecular bound state for positive values of the
 $s$-wave scattering length $a$ (the state bending down to negative
 infinity in the rel.\ motion spectrum in \figref{fig:icir}). If no
 c.m.-rel.\ coupling is present, as is the case for a harmonic
 confinement, the energy spectrum of the full two-body system is obtained
 by adding the energies of the rel.\ and c.m.\ motion, respectively. As
 a consequence, molecular bound states with c.m.\ excitation cross with
 states of unbound atom pairs denoted as trap states, i.\,e.\ states
 above $1.5\, \hbar \omega$ in the rel.\ motion plot. In case of a
 vanishing c.m.-rel.\ coupling ($W=0$) the states cross diabatically as
 indicated by the black dashed lines in \figref{fig:sketch} and is visible 
 in the inset in \figref{fig:icir} for $W=0$. 

 A coupling $W \ne 0$, 
 e.\,g.\ induced by an anharmonic trap, leads to an avoided crossing 
 and thus allows for
 an adiabatic transition (red solid line in \figref{fig:sketch}) of the
 trap state into a molecular state.
 This transfer into the bound state and thus a two-body recombination 
 is only possible, because the excess
 binding energy can be transferred into c.m.\ excitation energy due to
 the anharmonicity of the external confining potential. This redistribution of
 binding energy and kinetic energy is an inelastic process and thus
 these c.m.-rel.\ coupling resonances are denoted as \textit{inelastic}
 confinement-induced resonances \cite{cold:olsh98b}.  It was demonstrated
 in \cite{cold:sala13} that the adiabatic transfer into the molecular
 bound state and the resulting reduction of the interparticle distance
 by one order of magnitude can be performed fully coherently and
 controlled by tuning the scattering length using a magnetic Feshbach
 resonance.

 An alternative way to reach the resonance is to vary $d_{\rm ho}$, 
 i.\,e.\ the geometry of the external confinement which is feasible in
 Coulomb systems, e.\,g., for excitons in electrostatic traps. In
 contrast to atomic systems, where in the rel.\ motion spectrum of the
 $\delta$ pseudopotential only a single bound state exists,
 attractively interacting Coulomb systems like, e.\,g., excitons
 consisting of an electron and a hole, possess an infinite number of
 bound states that have the character of the ones of a hydrogen atom,
 i.\,e., they show a Rydberg series in free space approaching the
 continuum threshold. Inserting an exciton in a trap potential breaks
 the Rydberg series and leads to a smooth transition of bound to trap
 states. Hence the clear distinction of bound and trap states as it is
 present for the case of ultracold atoms is not possible.
 For repulsively interacting Coulomb systems, e.\,g., an
 electron-electron pair, there are no bound states present at all. A
 discretization is only induced by the trap potential. 
 Thus, the different structure of the spectra of QDs provokes the
 natural question whether ICIR occur in a QD and if an analog to the
 molecule formation in the atomic system can be observed.

%
\section{Computational approach for the energy spectrum of the coulomb system}
\label{sec:2}

 
 A system of two interacting particles with the absolute coordinates
 $\mathbf{r_1}$ and $\mathbf{r_2}$ is described by the Hamiltonian
\begin{align}
  H(\mathbf{r_1},\mathbf{r_2}) = \, & T_1(\mathbf{r_1}) +
  T_2(\mathbf{r_2})
  + V_1(\mathbf{r_1}) + V_2(\mathbf{r_2})  \nonumber \\
  & + U (\mathbf{|r_1-r_2|}) \, \textrm{,}
  \label{eq:HamABScoord}
\end{align}
 where $T_1$, $T_2$, $V_1$, and $V_2$ denote the kinetic and potential
 energies of the particles, respectively. The latter two represent the
 confinement of the QD. $U$ denotes the interparticle
 interaction. Since the Coulomb interaction
\begin{align}
  U_{\rm{Coul}}(r)=\pm \frac{1}{\epsilon_r} \frac{e^2}{4 \pi
    \epsilon_0 |\mathbf{r_1}-\mathbf{r_2}|} 
\end{align}
with $\epsilon_r$ as the dielectric constant depends solely on the interparticle distance, the Hamiltonian is
expressed in rel.\ and c.m.\ coordinates,
$\mathbf{r}=\mathbf{r_1}-\mathbf{r_2}$ and $\mathbf{R}=\frac{1}{2}
(\mathbf{r_1}+\mathbf{r_2})$, respectively,
\begin{align}
  \label{eq:HamRELCOM} 
  &H(\mathbf{r},\mathbf{R}) = H_{\mathrm{rel}}(\mathrm{r}) + H_{\mathrm{cm}}(\mathrm{R}) +  W(\mathbf{r},\mathbf{R}) \\
  &H_{\mathrm{rel}}(\mathrm{r})~\, = T_{\rm{rel}}(\mathbf{r}) +
  V_{\rm{rel}}(\mathbf{r}) + U_{\rm{Coul}} (r)  \\
  &H_{\mathrm{cm}}(\mathrm{R}) =    T_{\rm{cm}}(\mathbf{R}) + V_{\rm{cm}}(\mathbf{R}) \quad \textrm{.}
\end{align}
 Here, the potentials $V_{\rm{rel}}(\mathbf{r})$ and
 $V_{\rm{cm}}(\mathbf{R})$ represent the separable parts of the
 external potential, whereas the coupling term
 $W(\mathbf{r},\mathbf{R})$ includes all non-separable parts. Thus the
 Hamiltonians $H_{\mathrm{rel}}$ and $H_{\mathrm{cm}}$ correspond  
 effectively to single-particle systems and will occasionally be
 denoted in the following as single-particle Hamiltonians.

 For particles with identical mass confined to a harmonic potential the
 coupling term $W(\mathbf{r},\mathbf{R})$ vanishes and the Hamiltonian
 is composed of the rel.\ and c.m.\ motion of the two decoupled
 single-particle Hamiltonians, respectively.  While the solution of the
 c.m.\ motion is well-known for a harmonic confinement, due to the
 long-range behavior of the Coulomb interaction the rel.\ motion
 possesses only exact analytic solutions for certain energy levels
 using particular values of confinement and interaction
 strength \cite{qd:kest62,qd:que92,qd:star11}.

 In this work, the eigenenergies and wave functions of the stationary 
 Schr\"odinger equation with the Hamiltonian ($\ref{eq:HamRELCOM}$),
 \begin{align}
 H \, \ket{\Psi_i} = E_i \, \ket{\Psi_i} \, \textrm{,}
 \end{align}
 are calculated using an exact diagonalization approach \cite{cold:gris09,cold:gris11}. 
 Herein, the solutions of c.m.\ and rel.\ motion, 
 i.\,e.\ the decoupled parts of $H$, with their respective wave functions
 $\psi(\mathbf{R})$ and $\phi(\mathbf{r})$ are calculated separately. These 
 wave functions are expanded in B splines for the radial part and spherical 
 harmonics for the angular parts. The
 product states $\Phi_{\kappa} (\mathbf{R},\mathbf{r})=
 \psi_{i_{\kappa}}(\mathbf{R}) \, \phi_{j_{\kappa}}(\mathbf{r})$ form the
 basis of the solution of the full Hamiltonian. Thus, the full six-dimensional 
 wave functions of the two-particle system
 \begin{align}
 \Psi_i (\mathbf{R},\mathbf{r})= \sum_{\kappa} \, C_{i,\kappa} \, \Phi_{\kappa} (\mathbf{R},\mathbf{r})
 \label{eq:wf}
 \end{align}
 are constructed as superpositions of the $\Phi_{\kappa}$. 
 In order treat computationally efficiently 
 not only single wells (isolated quantum dots or ultracold atoms in a 
 single-well potential), but also quantum-dot molecules or ultracold atoms in 
 optical lattices, the basis functions are symmetry adapted to the eight 
 irreducible representations of the orthorhombic point group $D_{\rm{2h}}$ 
 $(A_{g},B_{1g},B_{2g},B_{3g},A_{u},B_{1u},B_{2u},B_{3u})$. This leads to a corresponding  
 block structure of the Hamiltonian matrix. 
 
 For locating ICIRs, a distinction of the different kinds of bound states for 
 the Coulomb system is necessary. 
 Similarly to the atomic system the mean interparticle distance $\overline{r}$
 can be chosen as a measure of the binding strength for the
 attractively interacting electron-hole system. 
 Thus, in order to characterize the Coulomb states, the  mean radial distance
 \begin{align}
   \overline{r} = \int_0^{\infty} \mathrm{d}r \, r\, \rho(r) 
   \label{eq:mean1}
 \end{align}
 is considered where
 \begin{align}
   \rho(r) = r^2 \, \int \, \mathrm{d}V_\mathbf{R} \, \mathrm{d}\Omega_\mathbf{r} \,
   | \Psi(\mathbf{r}, \mathbf{R}) |^2
   \label{eq:raddens}
 \end{align}
 is the radial pair density. The $\mathrm{d}V_\mathbf{R}$ 
 denotes the c.m.\ volume element and
 $\mathrm{d}\Omega_\mathbf{r}$ the angular volume element of the
 rel.\ motion.
 A state is regarded as a bound state, if $\overline{r}$ is of the same 
 order of magnitude as the effective Bohr radius $a_{\mu}=\epsilon_r / (\mu/m_o) \,
 a_B$, where $m_0$ is the rest mass of an electron and $a_B$
 the Bohr radius. If $\overline{r}$ is significantly larger than $a_{\mu}$, 
 one can consider this state as weakly bound and its behavior is expected to 
 be dominantly determined by the external trap potential. 
 
\begin{figure}[ht!]
  \begin{center}
    \includegraphics[width=0.45\textwidth]{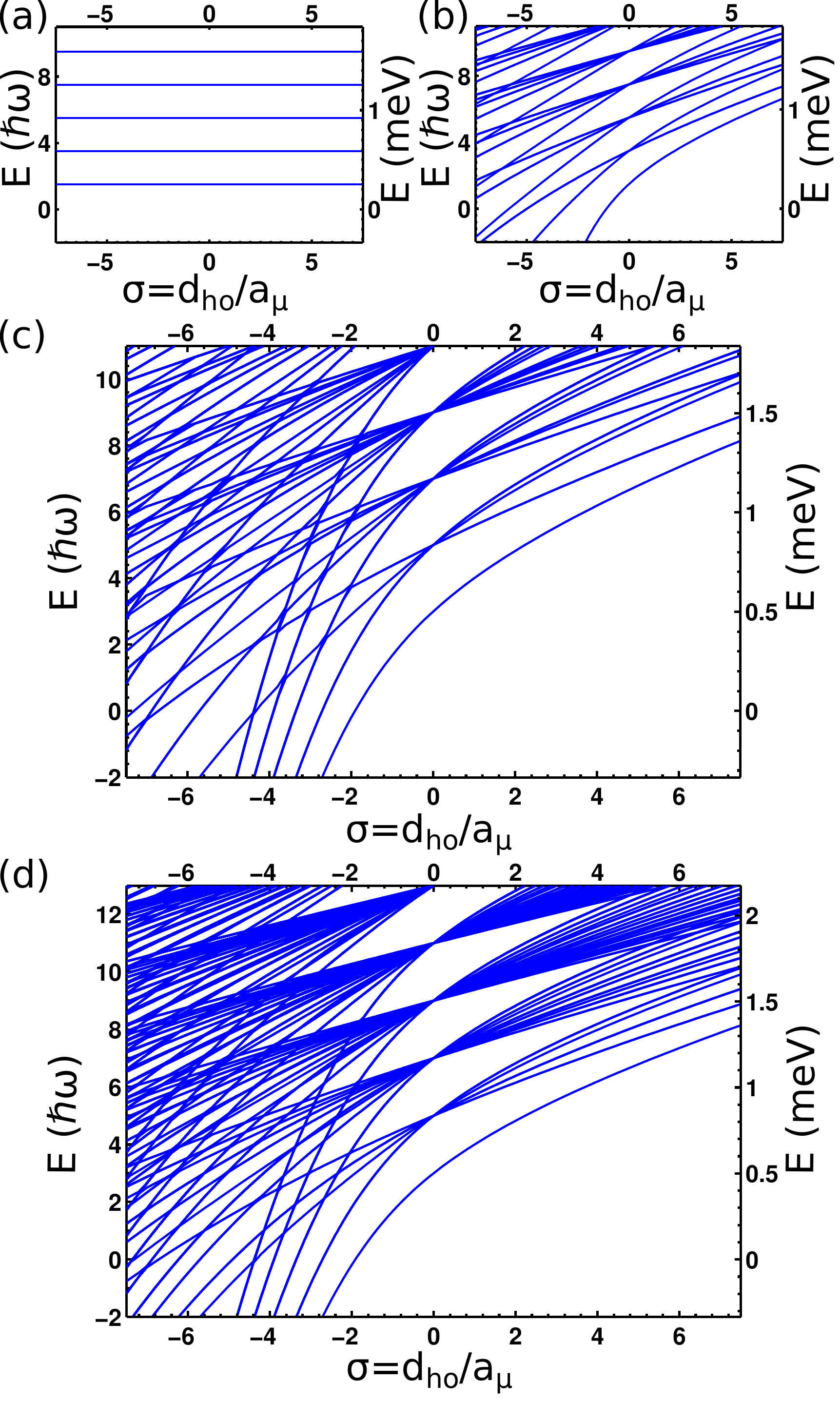}
    \caption{(Color online) Eigenenergy spectra for an exciton ($\sigma<0$) or an 
      electron pair ($\sigma>0$) with
      $m_{e,h}=0.067\,m_0$ confined in a harmonic potential with the
      dimensions $(d_x,d_y,d_z)=(31.6 , 31.6 , 31.6)\,\textrm{nm}\,$.
      The parameter $\sigma$ is modified by a change in $\epsilon_r$ 
      contained in $a_{\mu}$. 
      In (a) and (b) the $a_g$\ energy spectra of the rel.\ and
      c.m.\ motion, respectively, are shown. The $A_g$\ energy spectrum
      of the full Hamiltonian (\ref{eq:HamRELCOM}) for $W=0$ is shown
      when adopting only single-particle wave functions of $a_g$ symmetry (c)  
      or single-particle wave functions of all symmetries of the $D_{2h}$\ group (d). 
      }
   \label{fig:ElLoEMassesHarmVarQ}
  \end{center}
\end{figure}

 As introduction into the energy spectrum of Coulomb systems, 
 two Coulomb-interacting particles 
 of equal mass within the harmonic potential
 \begin{align}
 V_i(\mathbf{r_i}) = \sum_{j=x,y,z} \, V_j k_j^2 j^2
 \end{align}
 are considered. With the aid of the parameters $V_j$ and $k$ the
 potential depth and the size of the QDs are adjusted to the desired
 QD confinement.  A characteristic length scale of the potential is
 given by the harmonic-oscillator length $d_{j}=\sqrt{\hbar /(\mu
   \omega_j)}$ defined for each spatial direction $j=x,y,z$ with the
 harmonic-oscillator frequency $\omega_j=(V/\mu)^{1/2} \,k_j$ along 
 direction $j$. For an isotropic harmonic confinement for which $V=V_j$ and 
 $k=k_j$ and thus $d_{\mathrm{ho}}=d_{j}$ and $\omega=\omega_{j}$ applies, the
 energy spectrum of the two Coulomb-interacting particles is shown in
 \figref{fig:ElLoEMassesHarmVarQ}.

 Only wave functions of the total-symmetric 
 irreducible representation $a_g$\ of
 the decoupled single-particle spectra, shown in
 \figref{fig:ElLoEMassesHarmVarQ}(a) and (b), are adopted 
 when constructing the energy spectrum of the full six-dimensional 
 two-particle system in \figref{fig:ElLoEMassesHarmVarQ}(c). In
 \figref{fig:ElLoEMassesHarmVarQ}(d) the complete two-particle energy spectrum of 
 $A_g$\ symmetry is constructed by using the single-particle wave functions 
 of all symmetries. The orthorhombic D$_{2h}$ symmetry is only a 
 subgroup of the proper symmetry group of the here considered 
 spherical-symmetrical problem that results from the isotropic harmonic trap and  
 isotropic interparticle interaction. The orthorhombic 
 group is chosen since in this work mainly anisotropic systems are
 considered and, as discussed earlier, the code allows for the treatment of generically  
 orthorhombic problems like atoms in optical lattices or quantum-dot arrays.

 The parameter $\sigma=d / a_{\mu}$
 can be positive or negative depending 
 on whether the Coulomb interaction is attractive or repulsive, 
 representing an exciton system (negative $\sigma$) or an  
 electron pair (positive $\sigma$), respectively.
 The value of $\sigma$ depends on $\epsilon_r$ included in $a_{\mu}\,$. 
 A change in $\epsilon_r$ can be considered as a screening effect. 
 A variation of the
 interaction strength is a result of the modified properties of the
 environment, such as a modified charge density in the QD. However, a
 complete screening caused by $\epsilon_r\,$, which suppresses the
 long-range behavior of the Coulomb potential, is usually not obtained
 in semiconductors \cite{qd:czyc09}.

 The c.m.\ spectrum in \figref{fig:ElLoEMassesHarmVarQ} (a)
 reveals the well-known simple harmonic-oscillator solution
\begin{align}
    E_{\rm{ho}}=(2n + l + \frac{3}{2}) \hbar \omega \, \textrm{,}
\end{align}
 where the quantum numbers $n$ and $l$ are restricted to $0 \leq
 n,l$ ($l$ with only even numbers in $a_g$ symmetry). The energy 
 levels are not influenced by a change of the interaction
 strength. The relative-motion spectrum shown in
 \figref{fig:ElLoEMassesHarmVarQ}(b) varies with the interaction
 strength.  At $\sigma=0$, the Coulomb interaction is zero and the
 harmonic oscillator energies are revealed. For the even $a_g$\ symmetry,
 the ground state with $(n,l)=(0,0)$ appears. The next energy level at
 $\sigma=0$ comprises the states with quantum number $(n,l)$ equal to $(1,0)$ and $(0,2)$, 
 and thus, it is fourfold degenerate. 
 The degeneracy comes from the quantum number $m$ of the spherical
 harmonics with $-l \leq m \leq l$, but only even values occur due to the
 considered $a_g$ symmetry. A ninefold degenerate state with $(n,l)$ 
 corresponding to $(2,0)$,$(1,2)$ and $(0,4)$ follows,
 etc. For positive values of $\sigma$ the repulsive Coulomb
 interaction pushes the energy levels to higher values until they pass
 the continuum limit. For the attractive Coulomb interaction,
 i.\,e.\ $\sigma<0$, the states bend down in energy. Thereby the states
 with $l=0$ react more sensitively than higher $l>0$ states because of
 the presence of the centrifugal barrier for $l>0$. This leads to the
 occurrence of unavoided (real) crossings of states with different
 values of $l$. In the limit in which the trap depth approaches zero, 
 the attractive Coulomb interaction reproduces the Rydberg series again.

 The $A_g$\ energy spectrum of the full Hamiltonian
 (\ref{eq:HamABScoord}) shown in
 \figref{fig:ElLoEMassesHarmVarQ}(c) is only composed of $a_g$\ rel.\
 and c.m.\ states. Since the c.m.-rel.\ coupling vanishes for this
 system, the spectrum is obtained by adding the two single-particle
 spectra of the c.m.\ and rel.\ motion.  If all symmetries of $D_{2h}$
 from the rel.\ and c.m.\ motion are used for the construction of the
 full energy spectrum, as shown in \figref{fig:ElLoEMassesHarmVarQ}(d),
 some additional states appear that are (approximately) built from single-particles states
 of the same symmetry, e.\,g., $b_{1g} \otimes b_{1g} = A_g$ \cite{cold:gris11}. 
 The lowest energy levels still belong to the rel.\ and
 c.m.\ states in $a_g$\ symmetry as the comparison with the spectrum in
 \figref{fig:ElLoEMassesHarmVarQ}(c) shows.

 The $A_g$ symmetry is chosen, since it contains the ground state. If the coupling term
 $W(\mathbf{r},\mathbf{R})$ does not vanish, all symmetries should be
 included in the full energy-spectrum calculations since they contribute 
 to the composition of the full wave function $\Psi_i
 (\mathbf{R},\mathbf{r})$.

\section{C.m.-rel.\ coupling}
\label{sec:3}

 In order to simulate a more realistic single-well confinement of a QD, the 
 sextic potential \cite{cold:gris09}
\begin{align}
  V(\mathbf{r}) = \sum_{j=x,y,z} \, V_j \left( k_j^2 j^2 - \frac{1}{3}
    \, k_j^4 j^4 + \frac{2}{45} \, k_j^6 j^6 \right)
\end{align}
 is chosen. This single-well potential introduces the required anharmonicity for 
 c.m.-rel.\ coupling. In fact, the anharmonicity of this sextic potential is 
 relatively weak and thus the results discussed in this work are expected 
 to be a rather conservative estimate for the strength of the coupling 
 between c.m.\ and rel.\ motion as occurring in many real quantum dots, 
 especially those that are more appropriately modeled by a square-well 
 potential. 
 Since in the numerical approach the sextic potential is obtained by 
 an expansion of a $\sin^2$ potential up to the sixth order, 
 a parameter for the strength of the anharmonicity is the ratio 
 $V_j / (\hbar \omega_j)$. The deeper the potential, 
 i.\,e.\ the larger $V_j/(\hbar \omega_j) \gg 1$, the
 smaller the c.m.-rel.\ coupling (for fixed masses of the particles).
 
 Due to the anharmonicity of the confinement, the coupling term
 $W(\mathbf{r},\mathbf{R})$ in (\ref{eq:HamRELCOM}) does not vanish
 and is composed of non-separable parts of the form $r_j^n R_j^m$
 with $n,m \in \mathbb N \backslash \{ 0 \}$. If the two particles have
 equal effective masses, such as the electron pair, the
 non-separable part consists of a polynomial with even values of $n$
 and $m$, i.\,e.\ $r_j^2 R_j^2$, $r_j^2 R_j^4$ and $r_j^4 R_j^2$.  In the
 case of an exciton, where usually the electron and the hole have
 different effective masses, odd values of $n$ and $m$ also appear in
 $W(\mathbf{r},\mathbf{R})$, i.\,e.\ $r_j R_j$, $r_j R_j^3$, $r_j R_j^5$, $r_j^2 R_j^3$,
 $r_j^3 R_j$, and $r_j^5 R_j$.
 
 The matrix element
\begin{align}
  W_{\alpha,\beta}=
  \braket{\Psi^{(\alpha)}(\mathbf{r},\mathbf{R})|W(\mathbf{r},\mathbf{R})|\Psi^{(\beta)}(\mathbf{r},\mathbf{R})} 
  \label{eq:couplingW}
\end{align}
 defines the coupling strength between the states $\Psi^{(\alpha)}$ and
 $\Psi^{(\beta)}$. 
 In order to examine the states involved in a coupling at the resonance position, 
 their wave function densities are explored.

 \section{Results}
 \label{sec:4}

 In order to obtain ICIRs comparable with those in atomic
 systems \cite{cold:sala12,cold:sala13}, a strong anisotropic QD
 confinement is favored. In this work, inspired by the work on atomic systems, a
 cigar-shaped potential is chosen.  In this case the longitudinal and
 transversal excitations experience different trap strengths and thus
 have distinct extensions in the spatial directions providing a significant difference in the mean
 distance for a transition between different excitations of them. Since a very large anisotropy
 is numerically challenging for the adopted computational approach, the
 longitudinal size, i.\,e., the size in $x$ direction, is only chosen up to
 $\sqrt{10}$-times larger than the transversal sizes. Such  
 cigar-shaped potentials are encountered in quantum
 dashes \cite{qd:musi12}, nanorods \cite{qd:plan09}, and in the
 quasi 1D regime also known as quantum wires \cite{qd:tsuk92}.

 All energy spectra are calculated for the $A_g$\ symmetry and taking into account 
 single-particle wave functions of all symmetries. In the case of excitons    
 the particles usually possess different masses and are anyhow distinguishable. In 
 the case of an electron pair the particles are indistinguishable Fermions. Therefore, 
 states with $A_g$ symmetry in the spatial part correspond to spin-singlet 
 states in the case of electron pairs.

\subsection{Variation of the interaction strength}
\label{sec:enSp}
 In most cases, the effective mass of an electron is much smaller than
 the effective mass of the hole. However, first an electron and a hole in
 a sextic potential are considered where the hole mass is assumed to
 be equal to the effective electron mass, $m_h= m_e=0.067 \, m_0$. 
 Here, the c.m.-rel.\ coupling effects can only be caused 
 by the anharmonicity of the confinement, and thus, a clear
 distinction from other effects is possible. 
 
\begin{figure}[ht!]
  \begin{center}
   \includegraphics[width=0.45\textwidth]{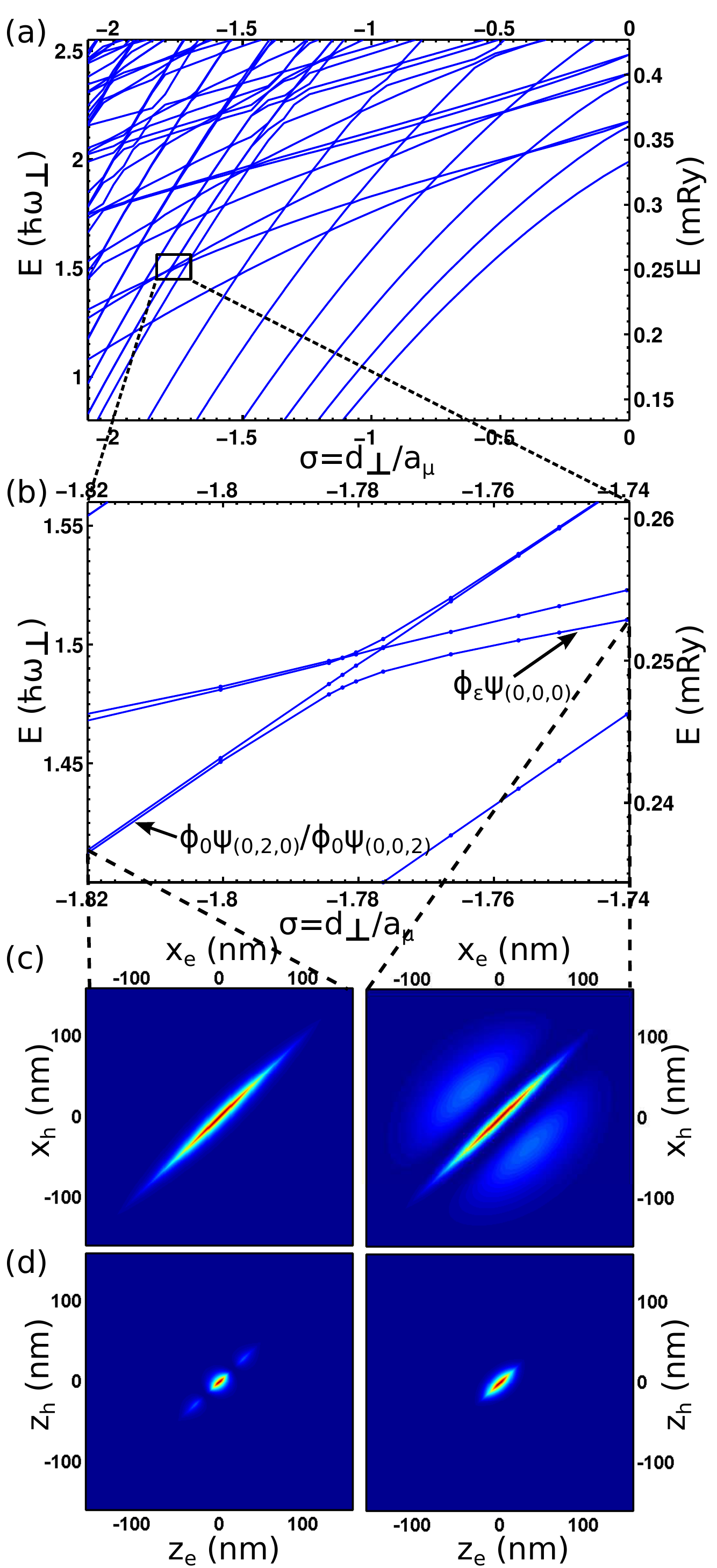}
   \caption{(Color online) 
   The $A_g$\ eigenenergy spectrum of the Hamiltonian (\ref{eq:HamRELCOM}) for an exciton 
   with $m_h=m_e=0.067\,m_0$ confined in a sextic potential with the dimensions 
   $(d_x,d_y,d_z)=(100 , 31.6 , 31.6)\,\textrm{nm}\,$ is shown in (a), 
   and the avoided crossing between coupled states, labeled with $\vartheta$, is magnified in (b). 
   Cuts through the wave-function density $|\Psi(x_e,y_e,z_e,x_h,y_h,z_h)|^2$ of the 
   coupled states along the $x$ direction in absolute coordinates, $x_h$ and $x_e$ 
   ($y_{e,h}=z_{e,h}=0$), in (c) and along the transversal $z$ direction in the 
   absolute coordinates, $z_h$ and $z_e$ ($x_{e,h}=y_{e,h}=0$), in (d) 
   confirm the c.m.-rel.\ coupling.} 
   \label{fig:ElLoEMassesSix10VarQ}
  \end{center}
 \end{figure}

 The fully coupled
 energy spectrum with variation of the Coulomb interaction strength is 
 shown in \figref{fig:ElLoEMassesSix10VarQ}(a).
 An avoided crossing can be identified in the framed box and is shown 
 magnified in \figref{fig:ElLoEMassesSix10VarQ}(b).  This avoided
 crossing, here labeled with $\vartheta$, with a full width at half maximum
 $\textrm{(FWHM)}\approx0.018$ has the position $d_{\perp}/a_{\mu}
 \approx -1.78$ with $a_{\mu} \approx 17.5\, \textrm{nm}$.
 
 The cuts through the wave-function density of the coupled states shown 
 in \figref{fig:ElLoEMassesSix10VarQ}(c) and (d) demonstrate a
 clear redistribution of the c.m.\ and rel.\ excitations. 
 For a further 
 illustration, a mapping of the states on familiar looking
 states of the c.m.\ and rel.\ motion from the Hamiltonian
 (\ref{eq:HamRELCOM}) can be performed. 
 Looking at the cuts of the wave-function density from 
 \figref{fig:ElLoEMassesSix10VarQ}(c), second-order c.m.\ excitations in 
 the transversal direction can be deduced from the three maxima of the density 
 along the diagonal with the unidirectional $z$ coordinates. Thus, the c.m.\ 
 motion can approximately be described by the notation $\psi_{\mathbf{n}} (\mathbf{R})$ with 
 $\mathbf{n}=(n_x,n_y,n_z)=(0,2,0)$ or $(0,0,2)$ inspired by 
 the quantum numbers $\mathbf{n}$ of the harmonic oscillator. Excitations in rel.\ motion 
 can be revealed by looking at the diagonals with counterpropagating coordinates, either in $x$ or in $z$ 
 direction, respectively. For the state in \figref{fig:ElLoEMassesSix10VarQ}(c), there is no rel.\ excitation, i.\,e., it is the ground state 
  $\phi_{0} (\mathbf{r})$. This bound state with a mean distance
  $\overline{r} \approx 16\,\textrm{nm}$ makes an adiabatic transition into a weaker bound state with rel.\
  excitation, but no c.m.\ excitation, i.\,e.\  $\psi_{(0,0,0)}(\mathbf{R}) \,
  \phi_{\varepsilon} (\mathbf{r})$, which has $\overline{r} \approx
  59\,\textrm{nm}$.
 This second state involved in the avoided crossing can be clearly identified 
 from the cuts of the wave-function density in \figref{fig:ElLoEMassesSix10VarQ}(d) 
 revealing two rel.\ 
 excitations in the longitudinal direction but no c.m.\ excitations.
 Both coupled states posses an even symmetry in the
  rel.\ and c.m.\ motions, respectively. This result is consistent with the
  condition that only states with the same symmetry can couple as is 
  required by the even powers of $\mathbf{r}$ and $\mathbf{R}$ in the
  coupling term $W(\mathbf{r},\mathbf{R})$ for equal-mass particles.
  This type of avoided crossing $\vartheta$ corresponds thus to 
  the ICIR found for atomic systems.
 
 Further ICIRs 
 between the transverse c.m.\ excited
 state and higher longitudinal rel.\ excited states occur. Here, a
 larger modification of the mean distance could be achieved. However, couplings between
 highly excited states are much weaker or even negligible due to the
 much smaller value of the integral in (\ref{eq:couplingW}) that expresses 
 the coupling strength.
 
\begin{figure}[ht!]
  \begin{center}
   \includegraphics[width=0.48\textwidth]{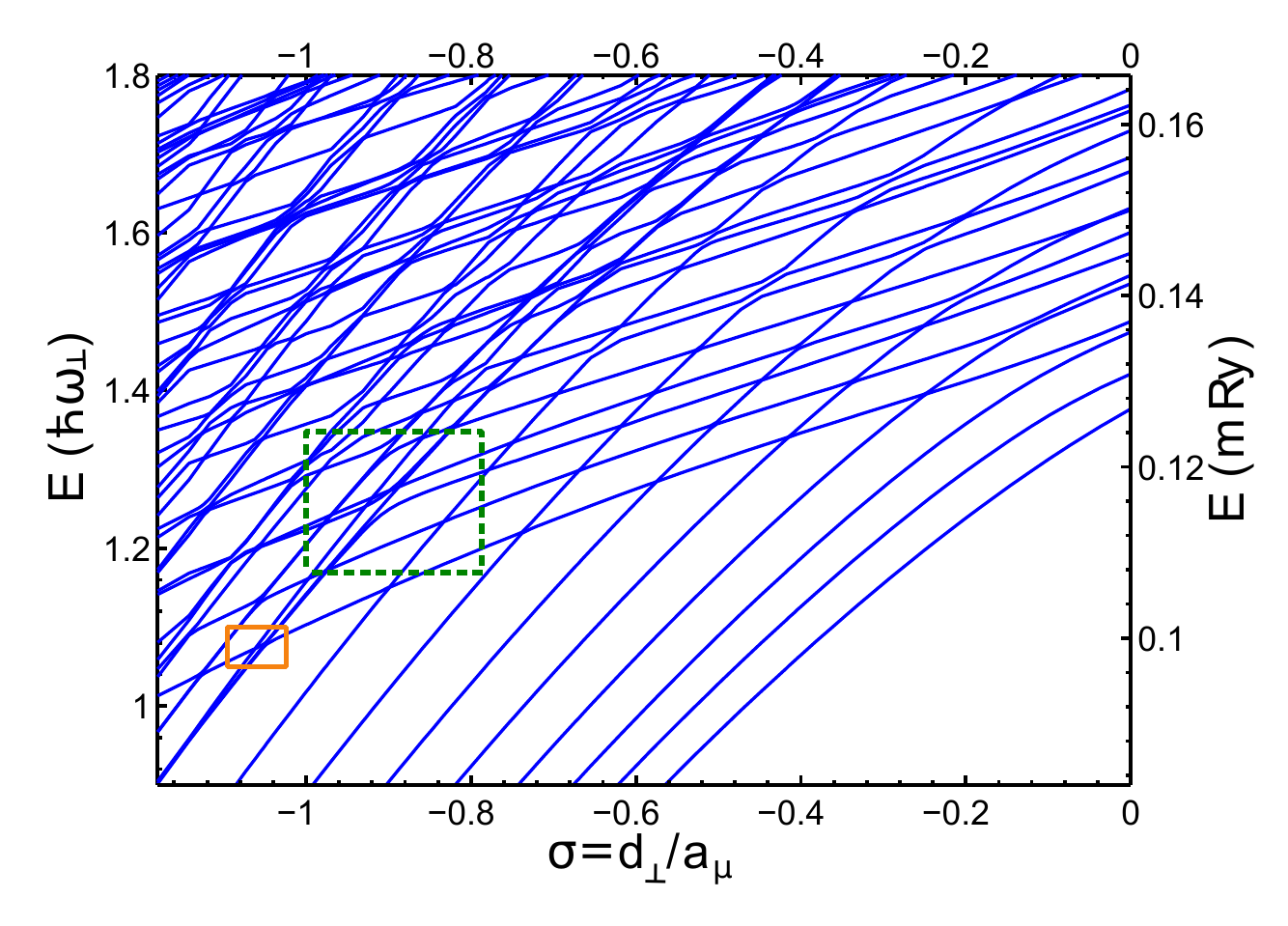}
   \caption{(Color online) 
   The $A_g$ eigenenergy spectrum of the Hamiltonian (\ref{eq:HamRELCOM}) for an exciton 
   with $m_h=10\,m_e=0.67\,m_0$ confined in a sextic potential with the dimensions
   $(d_x,d_y,d_z)=(100 , 31.6 , 31.6)\,\textrm{nm}$. The two squares indicate 
   avoided crossings, which are shown in detail in \figref{fig:ElLoDMassesSix10avcr1} (green dashed $\lambda_{\textrm{G}}$)
   and \figref{fig:ElLoDMassesSix10avcr2} (orange $\lambda_{\textrm{O}}$).}
   \label{fig:ElLoDMassesSix10VarQ}
  \end{center}
\end{figure}

 In the following, the particle-mass values are adjusted to realistic material properties of GaAs \cite{qd:chua09}.  
 The system of an electron and a hole with the mass $m_h=10 \, m_e=0.67 \, m_0$ is investigated. 
 Its energy spectrum is shown in \figref{fig:ElLoDMassesSix10VarQ}, 
 in which two avoided crossings are marked.
 
\begin{figure}[ht!] 
  \begin{center}
   \includegraphics[width=0.45\textwidth]{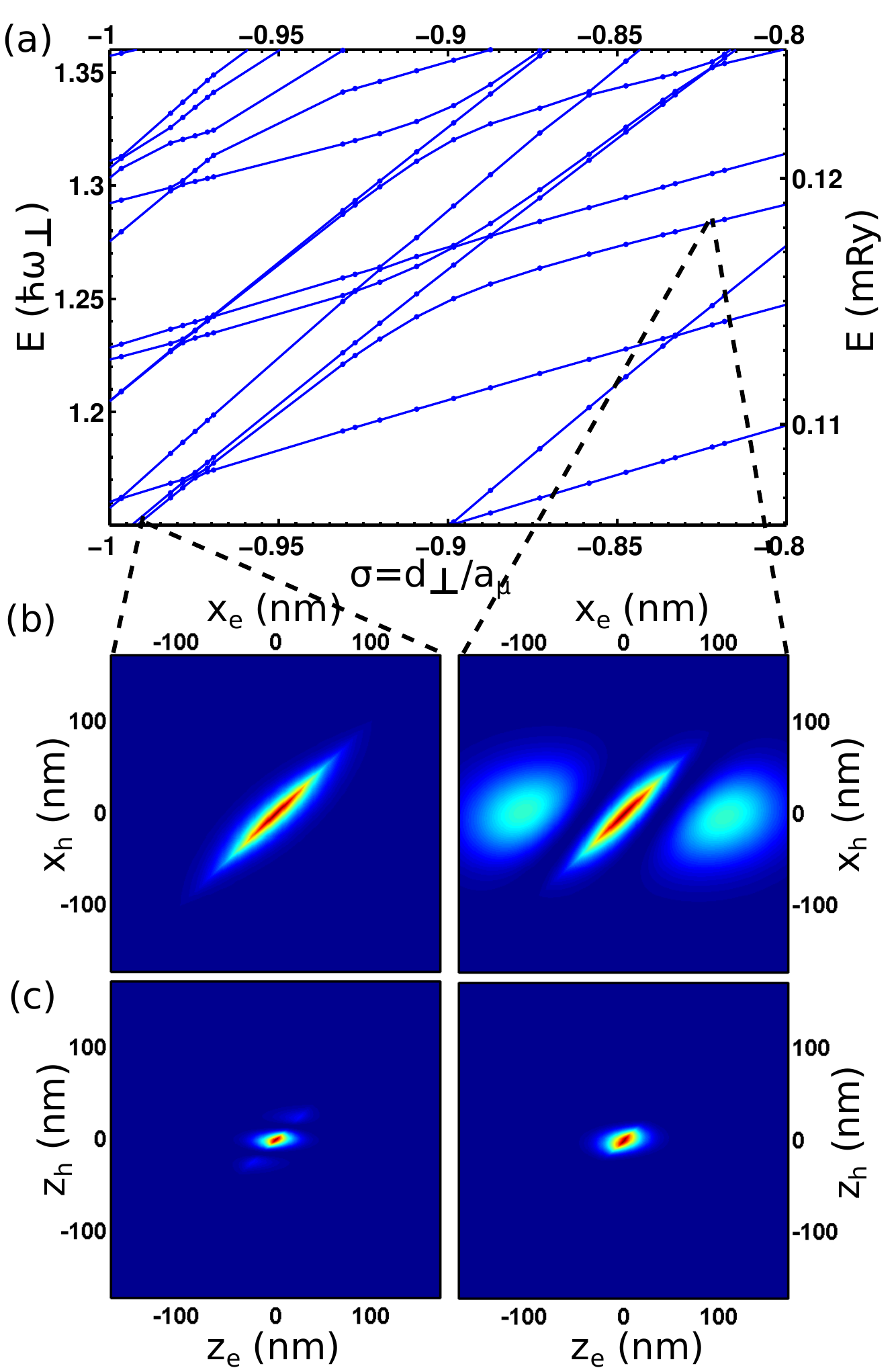}
   \caption{(Color online) 
   The avoided crossing $\lambda_{\textrm{G}}$ framed by the big green dashed square 
   in \figref{fig:ElLoDMassesSix10VarQ} magnified in (a). 
   Cuts through the wave function density $|\Psi(x_e,y_e,z_e,x_h,y_h,z_h)|^2$ of the 
   coupled states are shown along the absolute $x$ coordinates, $x_h$ and $x_e$ ($y_{e,h}=z_{e,h}=0$), 
   in (b) and along the absolute $z$-coordinates, $z_h$ and $z_e$ ($x_{e,h}=y_{e,h}=0$), 
   in (c).
   The c.m.-rel.\ resonance causes a significant change in distance 
   between the electron and the hole.}
   \label{fig:ElLoDMassesSix10avcr1}
  \end{center}
\end{figure}

\begin{figure}[ht!]
  \begin{center}
   \includegraphics[width=0.45\textwidth]{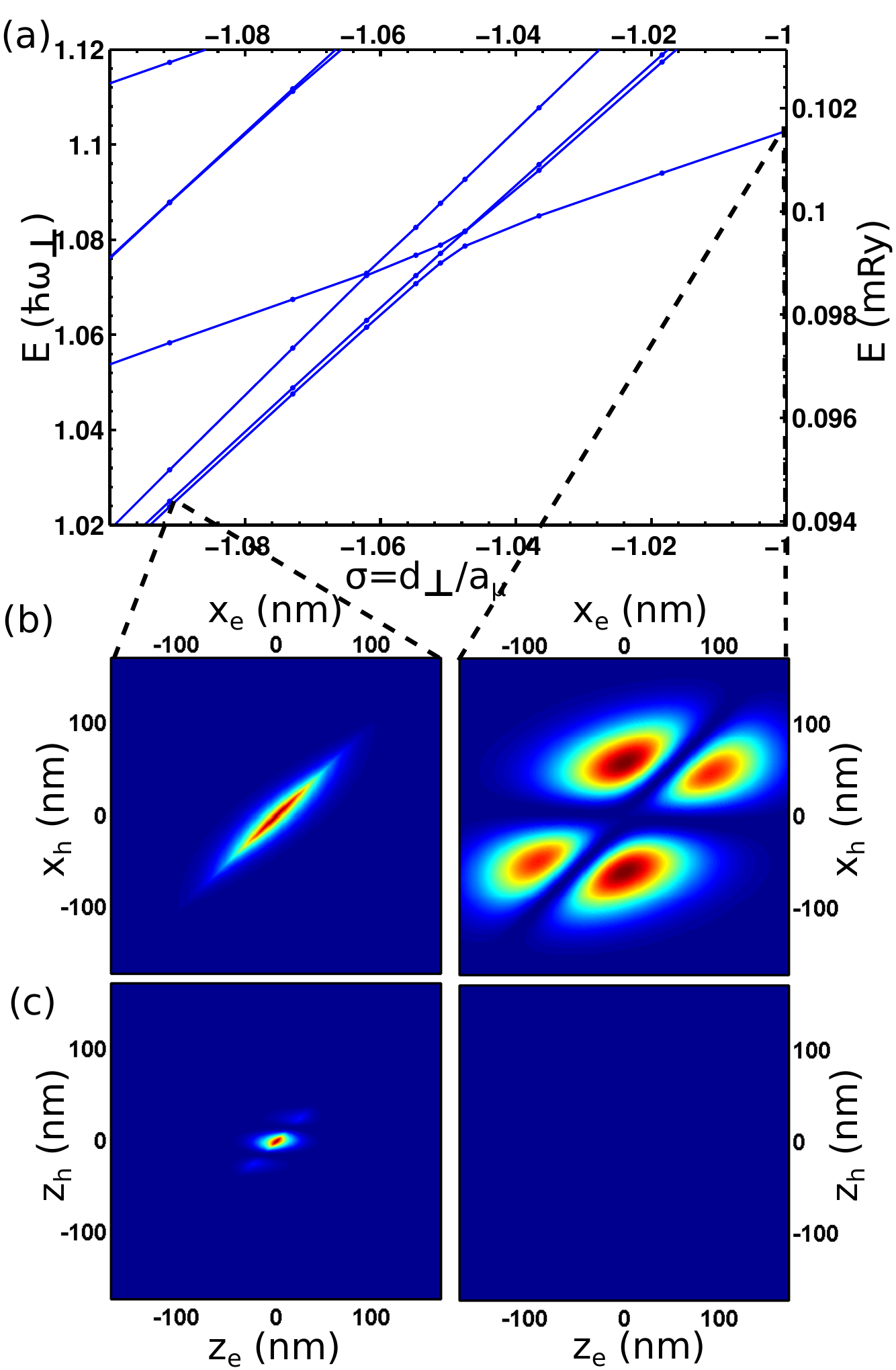}
   \caption{(Color online) 
   The avoided crossing $\lambda_{\textrm{O}}$ orange-framed 
   in \figref{fig:ElLoDMassesSix10VarQ} magnified in (a). 
   Cuts through the wave-function density $|\Psi(x_e,y_e,z_e,x_h,y_h,z_h)|^2$ of the coupled states are 
   shown along the absolute $x$ coordinates, $x_h$ and $x_e$ ($y_{e,h}=z_{e,h}=0$), in (b) 
   and along the absolute $z$ coordinates, $z_h$ and $z_e$ ($x_{e,h}=y_{e,h}=0$), in (c).
   The c.m.-rel.\ resonance causes a transition of a state with zero probability 
   for $d_{e,h}=0$ and a bound state.}
   \label{fig:ElLoDMassesSix10avcr2}
  \end{center}
\end{figure}

 The green dashed-framed avoided crossing presented in detail 
 in \figref{fig:ElLoDMassesSix10avcr1} 
 is labeled with $\lambda_{\textrm{G}}$ and has a 
 $\textrm{FWHM}\approx 0.059$ at position $d_{\perp}/a_{\mu}
 \approx -0.90$, where $a_{\mu} \approx 34.5\, \textrm{nm}$.
 Here, the weakly bound state with a mean distance of $\overline{r} \approx 80 \, \textrm{nm}$
 couples to a bound state with $\overline{r} \approx 24 \, \textrm{nm}$.
 A similar significant change in mean distance is found for the orange-framed avoided 
 crossing labeled with $\lambda_{\textrm{O}}$ and shown in detail in \figref{fig:ElLoDMassesSix10avcr2}.
 This crossing has a $\textrm{FWHM}\approx 0.006$ and is located at
 $d_{\perp}/a_{\mu} \approx -1.05$ with $a_{\mu} \approx 30
 \,\textrm{nm}$. Here, the same bound state as involved in 
 $\lambda_{\textrm{G}}$ goes over into a weak bound state with
 $\overline{r} \approx 65 \, \textrm{nm}$.  The interesting feature of
 $\lambda_{\textrm{O}}$ is the zero probability of the event that the
 two particles are located at the same place, i.\,e.\ the node at zero. 
 Hence, the recombination
 is blocked in the weakly bound state.  An adiabatic transition into this
 kind of states could have significant influence on photon emission.
 However, taking into account the coupling strength, which is discussed
 in more detail in Sec.\ \ref{sec:coupl}, the avoided crossing
 $\lambda_{\textrm{O}}$ is more difficult to be realized than the first avoided
 crossing $\lambda_{\textrm{G}}$.

 In general, the coupled states at both avoided crossings of \figref{fig:ElLoDMassesSix10VarQ} 
 are very different from the states of the system with equal-mass particles.
 The coupling terms that now contain also odd powers of $\mathbf{r}$ and
 $\mathbf{R}$ do not only cause avoided crossings between states, but also lead to 
 a general change of the shape of the wave functions. Thus, a mapping of the
 states to comparable product states of the c.m.\ and
 rel.\ motion appears not to be straightforward.
 
 Looking at unequal-mass particles in a harmonic confinement, avoided
 crossings can appear due to the non-vanishing coupling term $r_j R_j$.
 The coupling term again strongly influences the shape of the
 states. 
 Here, only avoided crossings similar to $\lambda_{\textrm{O}}$ can appear.

\begin{figure}[ht!]
  \begin{center}
   \includegraphics[width=0.45\textwidth]{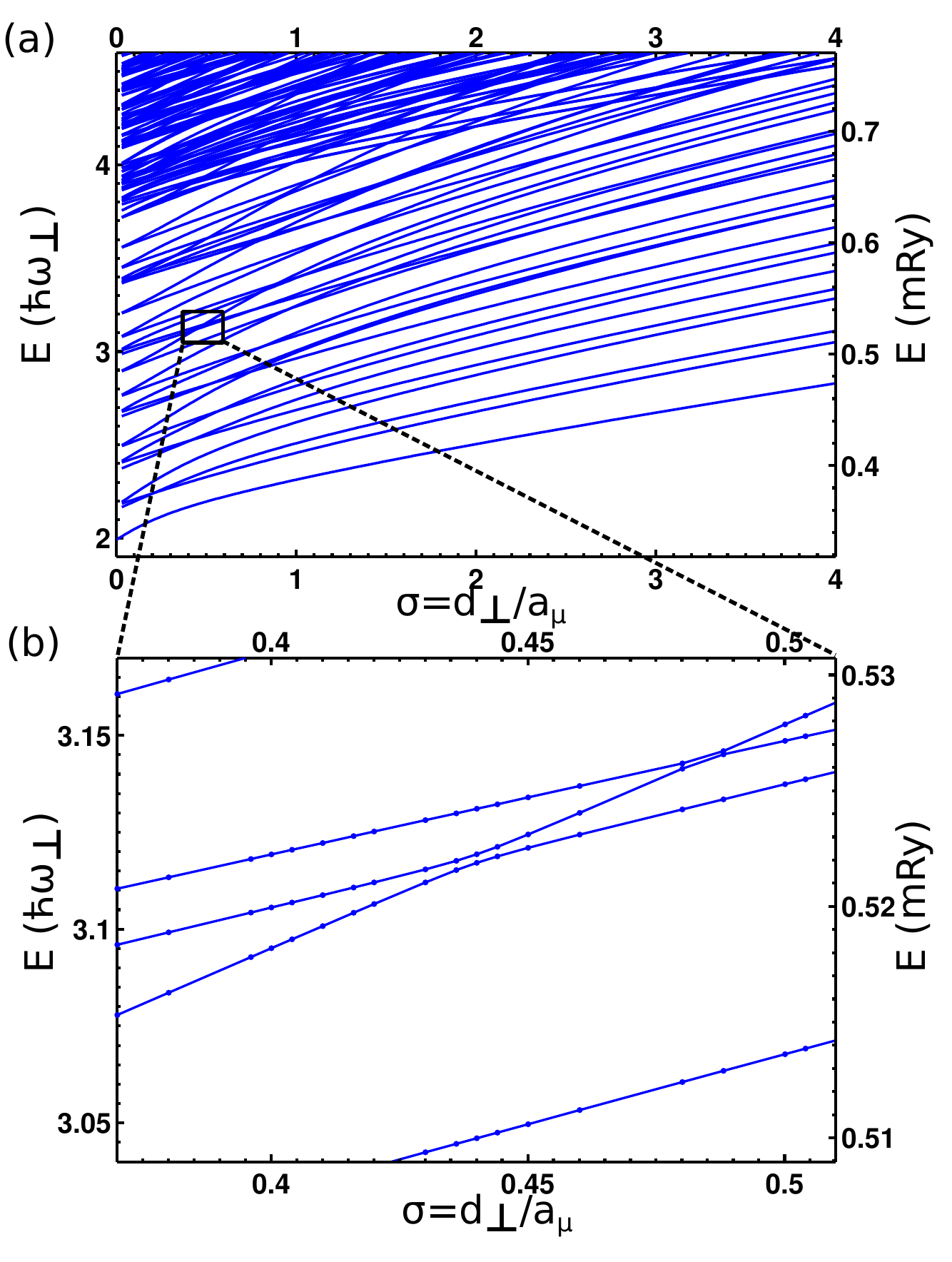}
   \caption{(Color online) 
   The $A_g$ eigenenergy spectrum of the Hamiltonian (\ref{eq:HamRELCOM}) for an 
   electron pair with $m_{e,1}=m_{e,2}=0.067\,m_0$ confined in 
   a sextic potential with the dimensions 
   $(d_x,d_y,d_z)=(100 , 31.6 , 31.6)\,\textrm{nm}\,$ 
   is shown in (a), and the avoided crossing between coupled states, label with $\xi$, is
   magnified in (b). Even for a repulsive 
   Coulomb interaction the c.m.-rel.\ coupling causes avoided crossings.}
   \label{fig:ElElEMassesSix10VarQ}
  \end{center}
\end{figure}

 In addition to the electron-hole system discussed above, the electron
 pair in a sextic potential is considered.  The energy spectrum shown
 in \figref{fig:ElElEMassesSix10VarQ}(a) reveals avoided crossings.
 Due to the repulsive Coulomb interaction and the confinement
 the states change only slightly in the interaction strength. Hence, only coupling between
 states with different longitudinal excitations is observed. One of these avoided crossings is labeled with $\xi$ and magnified in \figref{fig:ElElEMassesSix10VarQ}(b). Here, no significant change in the mean interparticle distance occurs. The first transversally excited states cross only with very highly lying longitudinally excited 
 states where due to the oscillatory behavior the coupling term becomes vanishing small. 
 For electrons in a stronger confinement a crossing between transversally excited and low 
 longitudinally excited states may, however, be realized. 
 Therefore, considering a QD array a transversal coupling to another QD due to the 
 tunnel effect is feasible. Consider the situation of a longitudinal state that is well 
 localized in a single QD and thus tunneling to a transversally coupled QD is suppressed, 
 since there is practically no electron density close to the tunnel barrier. If this 
 state is adiabatically transferred to the transversally excited state, this state 
 has a non-zero tunnel probability to the transversally coupled neighbor QD. Evidently, 
 an in-situ transfer form one state to the other may then be used as a switch that 
 induces a charge migration from one QD to another one.

\subsection{Analysis of the coupling strengths}
\label{sec:coupl}
 To discuss and compare the time-dependent behavior at the avoided crossings,  
 the Landau-Zener theory \cite{cold:zene32,cold:land32,cold:witt05} is used. 
 The probability of an adiabatic transition at the avoided crossing can be 
 estimated depending on the Landau-Zener velocity, i.\,e., the change in parameter $\sigma$ 
 per time $\textrm{d}\sigma/\textrm{d}t=v_{\textrm{LZ}}$. After calculating the linear 
 functions for the involved diabatic states and the coupling between them, the upper bound 
 of $v_{LZ}$ can be estimated at which the adiabatic transition is more likely than the diabatic transition.

 Applying the Landau-Zener theory to the avoided crossing $\vartheta$ of the electron-hole 
 system with equal-mass particles in \figref{fig:ElLoEMassesSix10VarQ}, 
 the velocity $v_{\textrm{LZ},\vartheta} \approx 0.9 \,\textrm{GHz}$ is found as 
 upper bound for an adiabatic transition. Regarding the required change in $\sigma$ 
 for passing this avoided crossing, $\textrm{d} \sigma \approx 0.1$, the transition 
 has to be performed within $\textrm{d}t_{\vartheta} \gtrsim 0.11 \, \textrm{ns}$. This 
 transition time is in the window of the exciton lifetime which is of the order 
 of $\tau \approx 1\, \textrm{ns}$ in GaAs \cite{qd:hoof97,qd:hwan73,qd:pere08}. 
 This result indicates the potential impact of an ICIRs in QDs.
 
 Considering the avoided crossings of the electron-hole system with $m_h=10\, m_e$ 
 in \figref{fig:ElLoDMassesSix10VarQ}, the upper bounds 
 $v_{\textrm{LZ},\lambda_{\textrm{G}}}\approx 2.9\, \textrm{GHz}$ for avoided 
 crossing $\lambda_{\textrm{G}}$ and $v_{\textrm{LZ},\lambda_{\textrm{O}}} \approx 34 \, \textrm{MHz}$ 
 for avoided crossing $\lambda_{\textrm{O}}$ are obtained. 
 Since $v_{\textrm{LZ},\lambda_{\textrm{G}}}$ is larger than $v_{\textrm{LZ},\vartheta}$, 
 different mass values of the particles enhance the coupling strength and an 
 adiabatic transition is dominant even for shorter time scales. Here, the transition 
 time $\textrm{d} t_{\lambda_{\textrm{G}}} \gtrsim 0.01 \,\textrm{ns}$ with a 
 required change $\textrm{d}\sigma \approx 0.15\,$ is obtained. 
 Contrary, observation of the avoided crossing of type $\lambda_{\textrm{O}}$ 
 is expected to be difficult due to the two orders of magnitude 
 smaller value of $v_{\textrm{LZ},\lambda_{\textrm{O}}}$ in comparison to 
 $v_{\textrm{LZ},\lambda_{\textrm{G}}}$. It has only the transition 
 time $\textrm{d} t_{\lambda_{\textrm{O}}} \gtrsim 0.9\, \textrm{ns}$ 
 with the change $\textrm{d}\sigma_{\lambda_{\textrm{O}}} \approx 0.03 \,$.
 
 For the avoided crossing $\xi$ of the electron pair from \figref{fig:ElElEMassesSix10VarQ}, 
 the velocity $v_{LZ,\xi}=0.15\, \textrm{GHz}$ is obtained.
 This velocity is considerably higher for transitions between two longitudinally  
 excited states of the electron pair. Here, a change $\textrm{d}\sigma_{\lambda_{\textrm{O}}} \approx 0.02 \,$ 
 with the transition time $\textrm{d} t_{\xi} \gtrsim 0.13\, \textrm{ns}$ has to be achieved.
 
 In general, the observation of an ICIR in experiments strongly depends on 
 the good resolution of the parameter $\sigma$. For this purpose, the 
 uncertainty of the dielectric constant\cite{qd:cham68}, of the 
 effective mass, and of the trap confinement must be kept low. 
 As mentioned earlier, it should on the other hand also be reminded that 
 the sextic potential chosen in the present simulations is only weakly 
 anharmonic. Trap potentials with higher degree of anharmonicity like 
 square-well potentials are expected to provide stronger couplings and 
 thus broader avoided crossings.

\subsection{Variation of the confinement}
\label{sec:conVar}
 The variation of the dielectric constant in a QD to change
 the Coulomb interaction strength between the particles is experimentally a complicated task. 
 A more suitable parameter for modification is the confinement of the QD. In 
 electrostatic QDs, a modification of the confinement
 is possible by a variation of the applied voltage \cite{qd:bedn03}. 
 The intrinsic problem arising in connection with electrostatic traps is 
 the exciton dissociation by applying an electric field. However, various 
 studies report electrostatic traps for indirect excitons in coupled 
 QDs \cite{qd:hube98,qd:hamm06,qd:high09,qd:schi13}. 
 Also for QDs fabricated by chemical and growth
 processes the size can be varied in a controlled way \cite{qd:saue05}. 
 Hence, it is appropriate to investigate a confinement variation
 for the electron-hole system. 

\begin{figure}[ht!]
  \begin{center}
   \includegraphics[width=0.45\textwidth]{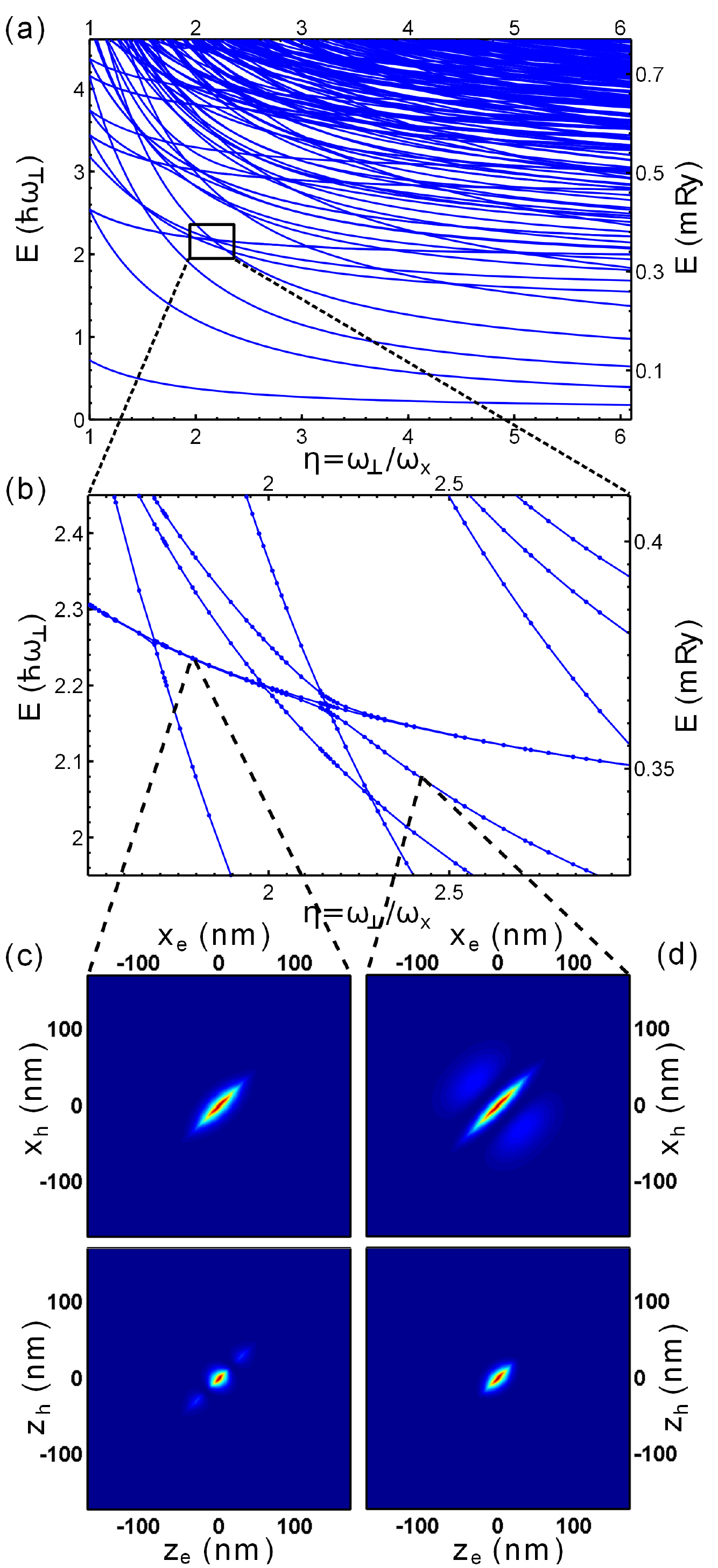}
   \caption{(Color online) 
   The $A_g$ eigenenergy spectrum of the Hamiltonian (\ref{eq:HamRELCOM}) for an exciton 
   with $m_h=m_e=0.067\,m_0$ confined in a sextic potential with the transversal 
   dimensions $d_{y,z}=31.6\,\textrm{nm}$ and dielectric constant $\epsilon_r=13.15$ giving a $d_{\perp}/a_{\mu} \approx -1.52$ 
   is shown (a), and an avoided crossing between coupled states, labeled with $\kappa$, is magnified in (b). 
   Cuts through the density $|\Psi(x_e,y_e,z_e,x_h,y_h,z_h)|^2$ of the 
   coupled states along the $x$ direction in absolute coordinates, $x_h$ and $x_e$ 
   ($y_{e,h}=z_{e,h}=0$), in (c) and along the transversal $z$ direction in the 
   absolute coordinates, $z_h$ and $z_e$ ($x_{e,h}=y_{e,h}=0$), in (d) are shown.}
   \label{fig:ElLoEMassesSix10VarInten}
  \end{center}
\end{figure}

 In \figref{fig:ElLoEMassesSix10VarInten}, the
 confinement is varied from an isotropic to a cigar-shaped potential with $x$ being the 
 longitudinal direction.
 The parameter $\eta=\omega_{\perp}/\omega_x$ indicates the anisotropy.  The energy levels bend down towards lower values the more anisotropic the confinement becomes. 
 An avoided crossing labeled with $\kappa$ can clearly be located at $\eta \approx 2.2\,$ and has a $\textrm{FWHM}\approx 0.120\,$. Here, a bound state with a small anisotropy of $\eta=1.9$ and a mean distance $\overline{r} \approx 18 \, \textrm{nm}$ couples to a weaker bound state with $\overline{r} \approx 45 \, \textrm{nm}$ at $\eta \approx 2.4 \, \textrm{nm}$. The densities (\figref{fig:ElLoEMassesSix10VarInten}(c) and (d)) of the avoided crossing  $\kappa$  reveal the redistribution of c.m.\ and rel.\ excitations, the typical characteristic of an ICIR. The change in mean distance is not as significant as in the case of the variation of the Coulomb interaction in \figref{fig:ElLoEMassesSix10VarQ}, since the anisotropy at the ICIR is smaller. However, 
 the position of the resonance may be tuned by varying the confinement strength in the transversal directions and thus leading to other changes in mean distance. 
 Moreover, for a system of particles with different masses a resonance with zero probability for 
 $d_{e,h}=0$ similar to the ICIR $\lambda_{O}$ can occur.
  
 Using the Landau-Zener theory, the upper bound $v_{\textrm{LZ},\kappa}\approx 0.8\, \textrm{GHz}$ for avoided 
 crossing $\kappa$ is found. This Landau-Zener velocity is similar to $v_{\textrm{LZ},\vartheta}$ obtained from 
 the variation of the Coulomb interaction. With the change $\textrm{d}\eta_{\kappa} \approx 0.5 \,$ in the 
 confinement variation, the transition time $\textrm{d} t_{\kappa} \gtrsim 0.63\, \textrm{ns}$ is found 
 which is still smaller than the usual exciton lifetimes \cite{qd:hoof97,qd:hwan73,qd:pere08}. 
 This feature makes ICIRs induced by a confinement variation accessible to an observation in experiments.

%
%
%
%
\section{Conclusion and outlook}
\label{sec:6}
 It is demonstrated that inelastic confinement-induced resonances,
 first described for trapped ultracold atoms \cite{cold:sala12,cold:sala13},
 occur in the electron-hole Coulomb systems confined in a
 QD. Furthermore, the c.m.-rel.\ coupling leads also to other types of
 avoided crossings as, e.\,g., in electron-electron systems.  In order to
 investigate a realistic QD, its confinement is approximated by an
 anharmonic and anisotropic potential.  For the electron-hole system,
 the ICIR causes a significant change in the interparticle distance, and thus can lead to
 a significant influence on the recombination rate. The change in
 distance grows with the anisotropy of the confinement, since couplings
 between weakly bound longitudinally excited states and bound transversally 
 excited states become possible. Thereby, the anharmonicity of the
 confinement induces the c.m.-rel.\ coupling. Hence, the coupling
 strength can be controlled by the confinement geometry.  Furthermore,
 looking at realistic exciton lifetimes a transition between states 
 at an ICIR appears to be in an experimentally feasible time window.

 An interesting application of the ICIR in Coulomb systems studied in 
 the present work could be a novel kind of single-photon sources on 
 demand based on excitons. In the case of excitons in self-organized QD 
 the \textit{in situ} variation of the confinement, i.\,e.\ the change 
 of the trap potential on a timescale shorter than the exciton 
 lifetime, is usually impossible. However, the realization of excitons 
 in electrostatic traps \cite{qd:hamm06,qd:high09,qd:schi13} 
 or laser-induced traps \cite{qd:hamm06b} overcomes this 
 limitation and thus allows for real-time reduction of the relative 
 mean distance between electron and hole at an ICIR. The enhanced 
 recombination probability leads then to a controlled photon 
 emission. However, in order to stabilize the exciton against 
 dissociation an additional electric field perpendicular to the plane 
 of electrostatic confinement is usually adopted. The influence of this 
 alignment of the exciton for the ICIR will be subject of a future 
 study. In the case of more than one exciton, the alignment leads to a 
 dipole-dipole interaction between the excitons \cite{qd:schi13}. Noteworthy, 
 the occurrence of ICIR in cold dipolar gases has recently 
 theoretically been demonstrated in \cite{cold:schu15}. 
 For electrons in QD arrays the ICIR may on the other hand be used 
 as a possible switch that allows to turn-on or -off the tunnel 
 current to a neighbor QD.
 
 Moreover, this work confirms the universal nature of ICIR as they are
 now proven to appear in the two very contrary systems of quantum dots and
 ultracold atoms.

\section*{Acknowledgments}
We are grateful to Alexander Stark for helpful first steps into this
research field. Furthermore we would like to thank the
\emph{Elsa-Neumann Stiftung}, \emph{Studienstiftung des deutschen Volkes},  
and \emph{Fonds der Chemischen Industrie} for financial 
support.


\begin{thebibliography}{42}%
\makeatletter
\providecommand \@ifxundefined [1]{%
 \@ifx{#1\undefined}
}%
\providecommand \@ifnum [1]{%
 \ifnum #1\expandafter \@firstoftwo
 \else \expandafter \@secondoftwo
 \fi
}%
\providecommand \@ifx [1]{%
 \ifx #1\expandafter \@firstoftwo
 \else \expandafter \@secondoftwo
 \fi
}%
\providecommand \natexlab [1]{#1}%
\providecommand \enquote  [1]{``#1''}%
\providecommand \bibnamefont  [1]{#1}%
\providecommand \bibfnamefont [1]{#1}%
\providecommand \citenamefont [1]{#1}%
\providecommand \href@noop [0]{\@secondoftwo}%
\providecommand \href [0]{\begingroup \@sanitize@url \@href}%
\providecommand \@href[1]{\@@startlink{#1}\@@href}%
\providecommand \@@href[1]{\endgroup#1\@@endlink}%
\providecommand \@sanitize@url [0]{\catcode `\\12\catcode `\$12\catcode
  `\&12\catcode `\#12\catcode `\^12\catcode `\_12\catcode `\%12\relax}%
\providecommand \@@startlink[1]{}%
\providecommand \@@endlink[0]{}
\providecommand \url  [0]{\begingroup\@sanitize@url \@url }%
\providecommand \@url [1]{\endgroup\@href {#1}{\urlprefix }}%
\providecommand \urlprefix  [0]{URL }%
\providecommand \Eprint [0]{\href }%
\providecommand \doibase [0]{http://dx.doi.org/}%
\providecommand \selectlanguage [0]{\@gobble}%
\providecommand \bibinfo  [0]{\@secondoftwo}%
\providecommand \bibfield  [0]{\@secondoftwo}%
\providecommand \translation [1]{[#1]}%
\providecommand \BibitemOpen [0]{}%
\providecommand \bibitemStop [0]{}%
\providecommand \bibitemNoStop [0]{.\EOS\space}%
\providecommand \EOS [0]{\spacefactor3000\relax}%
\providecommand \BibitemShut  [1]{\csname bibitem#1\endcsname}%
\let\auto@bib@innerbib\@empty
\bibitem [{\citenamefont {Bera}\ \emph {et~al.}(2010)\citenamefont {Bera},
  \citenamefont {Qian}, \citenamefont {Teng-Kuan},\ and\ \citenamefont
  {Holloway}}]{qd:bera10}%
  \BibitemOpen
  \bibfield  {author} {\bibinfo {author} {\bibfnamefont {D.}~\bibnamefont
  {Bera}}, \bibinfo {author} {\bibfnamefont {L.}~\bibnamefont {Qian}}, \bibinfo
  {author} {\bibnamefont {Teng-Kuan}}, \ and\ \bibinfo {author} {\bibfnamefont
  {P.~H.}\ \bibnamefont {Holloway}},\ }\href@noop {} {\bibfield  {journal}
  {\bibinfo  {journal} {Materials}\ }\textbf {\bibinfo {volume} {3}},\ \bibinfo
  {pages} {2260} (\bibinfo {year} {2010})}\BibitemShut {NoStop}%
\bibitem [{\citenamefont {Eisaman}\ \emph {et~al.}(2011)\citenamefont
  {Eisaman}, \citenamefont {Fan}, \citenamefont {Migdall},\ and\ \citenamefont
  {Polyakov}}]{qd:eisa11}%
  \BibitemOpen
  \bibfield  {author} {\bibinfo {author} {\bibfnamefont {M.~D.}\ \bibnamefont
  {Eisaman}}, \bibinfo {author} {\bibfnamefont {J.}~\bibnamefont {Fan}},
  \bibinfo {author} {\bibfnamefont {A.}~\bibnamefont {Migdall}}, \ and\
  \bibinfo {author} {\bibfnamefont {S.~V.}\ \bibnamefont {Polyakov}},\
  }\href@noop {} {\bibfield  {journal} {\bibinfo  {journal} {Rev. Sci.
  Instrum.}\ }\textbf {\bibinfo {volume} {82}},\ \bibinfo {pages} {071101}
  (\bibinfo {year} {2011})}\BibitemShut {NoStop}%
\bibitem [{\citenamefont {Claudon}\ \emph {et~al.}(2010)\citenamefont
  {Claudon}, \citenamefont {Bleuse}, \citenamefont {Malik}, \citenamefont
  {Bazin}, \citenamefont {Jaffrennou}, \citenamefont {Gregersen}, \citenamefont
  {Sauvan}, \citenamefont {Lalanne},\ and\ \citenamefont
  {G\'{e}rard}}]{qd:clau10}%
  \BibitemOpen
  \bibfield  {author} {\bibinfo {author} {\bibfnamefont {J.}~\bibnamefont
  {Claudon}}, \bibinfo {author} {\bibfnamefont {J.}~\bibnamefont {Bleuse}},
  \bibinfo {author} {\bibfnamefont {N.~S.}\ \bibnamefont {Malik}}, \bibinfo
  {author} {\bibfnamefont {M.}~\bibnamefont {Bazin}}, \bibinfo {author}
  {\bibfnamefont {P.}~\bibnamefont {Jaffrennou}}, \bibinfo {author}
  {\bibfnamefont {N.}~\bibnamefont {Gregersen}}, \bibinfo {author}
  {\bibfnamefont {C.}~\bibnamefont {Sauvan}}, \bibinfo {author} {\bibfnamefont
  {P.}~\bibnamefont {Lalanne}}, \ and\ \bibinfo {author} {\bibfnamefont
  {J.-M.}\ \bibnamefont {G\'{e}rard}},\ }\href@noop {} {\bibfield  {journal}
  {\bibinfo  {journal} {Nature Photonics}\ }\textbf {\bibinfo {volume} {4}},\
  \bibinfo {pages} {174} (\bibinfo {year} {2010})}\BibitemShut {NoStop}%
\bibitem [{\citenamefont {Devoret}\ and\ \citenamefont
  {Schoelkopf}(2000)}]{qd:devo00}%
  \BibitemOpen
  \bibfield  {author} {\bibinfo {author} {\bibfnamefont {M.~H.}\ \bibnamefont
  {Devoret}}\ and\ \bibinfo {author} {\bibfnamefont {R.~J.}\ \bibnamefont
  {Schoelkopf}},\ }\href@noop {} {\bibfield  {journal} {\bibinfo  {journal}
  {Nature}\ }\textbf {\bibinfo {volume} {406}},\ \bibinfo {pages} {1039}
  (\bibinfo {year} {2000})}\BibitemShut {NoStop}%
\bibitem [{\citenamefont {Stampfer}\ \emph {et~al.}(2008)\citenamefont
  {Stampfer}, \citenamefont {Schurtenberger}, \citenamefont {Molitor},
  \citenamefont {G\"uttinger}, \citenamefont {Ihn},\ and\ \citenamefont
  {Ensslin}}]{qd:stam08}%
  \BibitemOpen
  \bibfield  {author} {\bibinfo {author} {\bibfnamefont {C.}~\bibnamefont
  {Stampfer}}, \bibinfo {author} {\bibfnamefont {E.}~\bibnamefont
  {Schurtenberger}}, \bibinfo {author} {\bibfnamefont {F.}~\bibnamefont
  {Molitor}}, \bibinfo {author} {\bibfnamefont {J.}~\bibnamefont
  {G\"uttinger}}, \bibinfo {author} {\bibfnamefont {T.}~\bibnamefont {Ihn}}, \
  and\ \bibinfo {author} {\bibfnamefont {K.}~\bibnamefont {Ensslin}},\
  }\href@noop {} {\bibfield  {journal} {\bibinfo  {journal} {Nano Lett.}\
  }\textbf {\bibinfo {volume} {8}},\ \bibinfo {pages} {2378} (\bibinfo {year}
  {2008})}\BibitemShut {NoStop}%
\bibitem [{\citenamefont {Alivisatos}(1996)}]{qd:aliv96}%
  \BibitemOpen
  \bibfield  {author} {\bibinfo {author} {\bibfnamefont {A.~P.}\ \bibnamefont
  {Alivisatos}},\ }\href@noop {} {\bibfield  {journal} {\bibinfo  {journal}
  {Science}\ }\textbf {\bibinfo {volume} {271}},\ \bibinfo {pages} {5251}
  (\bibinfo {year} {1996})}\BibitemShut {NoStop}%
\bibitem [{\citenamefont {\"Unl\"u}\ \emph {et~al.}(2006)\citenamefont
  {\"Unl\"u}, \citenamefont {Karabulut},\ and\ \citenamefont
  {\c{S}afak}}]{qd:unlu06}%
  \BibitemOpen
  \bibfield  {author} {\bibinfo {author} {\bibfnamefont {S.}~\bibnamefont
  {\"Unl\"u}}, \bibinfo {author} {\bibfnamefont {I.}~\bibnamefont {Karabulut}},
  \ and\ \bibinfo {author} {\bibfnamefont {H.}~\bibnamefont {\c{S}afak}},\
  }\href@noop {} {\bibfield  {journal} {\bibinfo  {journal} {Physica E}\
  }\textbf {\bibinfo {volume} {33}},\ \bibinfo {pages} {319} (\bibinfo {year}
  {2006})}\BibitemShut {NoStop}%
\bibitem [{\citenamefont {Brey}\ \emph {et~al.}(1989)\citenamefont {Brey},
  \citenamefont {Johnson},\ and\ \citenamefont {Halperin}}]{qd:brey89}%
  \BibitemOpen
  \bibfield  {author} {\bibinfo {author} {\bibfnamefont {L.}~\bibnamefont
  {Brey}}, \bibinfo {author} {\bibfnamefont {N.~F.}\ \bibnamefont {Johnson}}, \
  and\ \bibinfo {author} {\bibfnamefont {B.~I.}\ \bibnamefont {Halperin}},\
  }\href@noop {} {\bibfield  {journal} {\bibinfo  {journal} {Phys.\,Rev.\,B}\
  }\textbf {\bibinfo {volume} {40}},\ \bibinfo {pages} {10647(R)} (\bibinfo
  {year} {1989})}\BibitemShut {NoStop}%
\bibitem [{\citenamefont {Kumar}\ \emph {et~al.}(1990)\citenamefont {Kumar},
  \citenamefont {Laux},\ and\ \citenamefont {Stern}}]{qd:kuma90}%
  \BibitemOpen
  \bibfield  {author} {\bibinfo {author} {\bibfnamefont {A.}~\bibnamefont
  {Kumar}}, \bibinfo {author} {\bibfnamefont {S.~E.}\ \bibnamefont {Laux}}, \
  and\ \bibinfo {author} {\bibfnamefont {F.}~\bibnamefont {Stern}},\
  }\href@noop {} {\bibfield  {journal} {\bibinfo  {journal} {Phys.\,Rev.\,B}\
  }\textbf {\bibinfo {volume} {42}},\ \bibinfo {pages} {5166} (\bibinfo {year}
  {1990})}\BibitemShut {NoStop}%
\bibitem [{\citenamefont {Que}(1992)}]{qd:que92}%
  \BibitemOpen
  \bibfield  {author} {\bibinfo {author} {\bibfnamefont {W.}~\bibnamefont
  {Que}},\ }\href@noop {} {\bibfield  {journal} {\bibinfo  {journal}
  {Phys.\,Rev.\,B}\ }\textbf {\bibinfo {volume} {45}},\ \bibinfo {pages}
  {11036} (\bibinfo {year} {1992})}\BibitemShut {NoStop}%
\bibitem [{\citenamefont {Sala}\ \emph {et~al.}(2012)\citenamefont {Sala},
  \citenamefont {Schneider},\ and\ \citenamefont {Saenz}}]{cold:sala12}%
  \BibitemOpen
  \bibfield  {author} {\bibinfo {author} {\bibfnamefont {S.}~\bibnamefont
  {Sala}}, \bibinfo {author} {\bibfnamefont {P.-I.}\ \bibnamefont {Schneider}},
  \ and\ \bibinfo {author} {\bibfnamefont {A.}~\bibnamefont {Saenz}},\
  }\href@noop {} {\bibfield  {journal} {\bibinfo  {journal}
  {Phys.\,Rev.\,Lett.}\ }\textbf {\bibinfo {volume} {109}},\ \bibinfo {pages}
  {073201} (\bibinfo {year} {2012})}\BibitemShut {NoStop}%
\bibitem [{\citenamefont {Sala}\ \emph {et~al.}(2013)\citenamefont {Sala},
  \citenamefont {Z{\"u}rn}, \citenamefont {Lompe}, \citenamefont {Wenz},
  \citenamefont {Murmann}, \citenamefont {Serwane}, \citenamefont {Jochim},\
  and\ \citenamefont {Saenz}}]{cold:sala13}%
  \BibitemOpen
  \bibfield  {author} {\bibinfo {author} {\bibfnamefont {S.}~\bibnamefont
  {Sala}}, \bibinfo {author} {\bibfnamefont {G.}~\bibnamefont {Z{\"u}rn}},
  \bibinfo {author} {\bibfnamefont {T.}~\bibnamefont {Lompe}}, \bibinfo
  {author} {\bibfnamefont {A.~N.}~\bibnamefont {Wenz}}, \bibinfo {author}
  {\bibfnamefont {S.}~\bibnamefont {Murmann}}, \bibinfo {author} {\bibfnamefont
  {F.}~\bibnamefont {Serwane}}, \bibinfo {author} {\bibfnamefont
  {S.}~\bibnamefont {Jochim}}, \ and\ \bibinfo {author} {\bibfnamefont
  {A.}~\bibnamefont {Saenz}},\ }\href@noop {} {\bibfield  {journal} {\bibinfo
  {journal} {Phys.\,Rev.\,Lett.}\ }\textbf {\bibinfo {volume} {110}},\ \bibinfo
  {pages} {203202} (\bibinfo {year} {2013})}\BibitemShut {NoStop}%
\bibitem [{col()}]{cold:olsh98b}%
  \BibitemOpen
  \href@noop {} {}\bibinfo {note} {There exists also a theory of
  \textit{elastic} CIRs for atomic systems introduced by M.\ Olshanii
  [\textit{Phys. Rev. Lett.} \textbf{81}, 938 (1998)]. They are based only on
  the rel.\ motion within a purely harmonic but strongly anisotropic, quasi
  one-dimensional confinement.}\BibitemShut {Stop}%
\bibitem [{\citenamefont {Christen}\ \emph {et~al.}(1984)\citenamefont
  {Christen}, \citenamefont {Bimberg}, \citenamefont {Steckenborn},\ and\
  \citenamefont {Weiman}}]{qd:chri84}%
  \BibitemOpen
  \bibfield  {author} {\bibinfo {author} {\bibfnamefont {J.}~\bibnamefont
  {Christen}}, \bibinfo {author} {\bibfnamefont {D.}~\bibnamefont {Bimberg}},
  \bibinfo {author} {\bibfnamefont {A.}~\bibnamefont {Steckenborn}}, \ and\
  \bibinfo {author} {\bibfnamefont {G.}~\bibnamefont {Weiman}},\ }\href@noop {}
  {\bibfield  {journal} {\bibinfo  {journal} {Appl. Phys. Lett.}\ }\textbf
  {\bibinfo {volume} {44}},\ \bibinfo {pages} {84} (\bibinfo {year}
  {1984})}\BibitemShut {NoStop}%
\bibitem [{\citenamefont {Haller}\ \emph {et~al.}(2010)\citenamefont {Haller},
  \citenamefont {Mark}, \citenamefont {Hart}, \citenamefont {Danzl},
  \citenamefont {Reich{\-}s{\"o}l{\-}lner}, \citenamefont {Melezhik},
  \citenamefont {Schmelcher},\ and\ \citenamefont {N{\"a}gerl}}]{cold:hall10b}%
  \BibitemOpen
  \bibfield  {author} {\bibinfo {author} {\bibfnamefont {E.}~\bibnamefont
  {Haller}}, \bibinfo {author} {\bibfnamefont {M.~J.}\ \bibnamefont {Mark}},
  \bibinfo {author} {\bibfnamefont {R.}~\bibnamefont {Hart}}, \bibinfo {author}
  {\bibfnamefont {J.~G.}\ \bibnamefont {Danzl}}, \bibinfo {author}
  {\bibfnamefont {L.}~\bibnamefont {Reich{\-}s{\"o}l{\-}lner}}, \bibinfo
  {author} {\bibfnamefont {V.}~\bibnamefont {Melezhik}}, \bibinfo {author}
  {\bibfnamefont {P.}~\bibnamefont {Schmelcher}}, \ and\ \bibinfo {author}
  {\bibfnamefont {H.-C.}\ \bibnamefont {N{\"a}gerl}},\ }\href@noop {}
  {\bibfield  {journal} {\bibinfo  {journal} {Phys.\,Rev.\,Lett.}\ }\textbf
  {\bibinfo {volume} {104}},\ \bibinfo {pages} {153203} (\bibinfo {year}
  {2010})}\BibitemShut {NoStop}%
\bibitem [{\citenamefont {Chin}\ \emph {et~al.}(2010)\citenamefont {Chin},
  \citenamefont {Grimm}, \citenamefont {Julienne},\ and\ \citenamefont
  {Tiesinga}}]{cold:chin10}%
  \BibitemOpen
  \bibfield  {author} {\bibinfo {author} {\bibfnamefont {C.}~\bibnamefont
  {Chin}}, \bibinfo {author} {\bibfnamefont {R.}~\bibnamefont {Grimm}},
  \bibinfo {author} {\bibfnamefont {P.}~\bibnamefont {Julienne}}, \ and\
  \bibinfo {author} {\bibfnamefont {E.}~\bibnamefont {Tiesinga}},\ }\href@noop
  {} {\bibfield  {journal} {\bibinfo  {journal} {Rev.\,Mod.\,Phys.}\ }\textbf
  {\bibinfo {volume} {82}},\ \bibinfo {pages} {1225} (\bibinfo {year}
  {2010})}\BibitemShut {NoStop}%
\bibitem [{\citenamefont {Busch}\ \emph {et~al.}(1998)\citenamefont {Busch},
  \citenamefont {Englert}, \citenamefont {Rzazewski},\ and\ \citenamefont
  {Wilkens}}]{cold:busc98}%
  \BibitemOpen
  \bibfield  {author} {\bibinfo {author} {\bibfnamefont {T.}~\bibnamefont
  {Busch}}, \bibinfo {author} {\bibfnamefont {B.-G.}\ \bibnamefont {Englert}},
  \bibinfo {author} {\bibfnamefont {K.}~\bibnamefont {Rzazewski}}, \ and\
  \bibinfo {author} {\bibfnamefont {M.}~\bibnamefont {Wilkens}},\ }\href@noop
  {} {\bibfield  {journal} {\bibinfo  {journal} {Found.\,Phys.}\ }\textbf
  {\bibinfo {volume} {28}},\ \bibinfo {pages} {549} (\bibinfo {year}
  {1998})}\BibitemShut {NoStop}%
\bibitem [{\citenamefont {Idziaszek}\ and\ \citenamefont
  {Calarco}(2006)}]{cold:idzi06}%
  \BibitemOpen
  \bibfield  {author} {\bibinfo {author} {\bibfnamefont {Z.}~\bibnamefont
  {Idziaszek}}\ and\ \bibinfo {author} {\bibfnamefont {T.}~\bibnamefont
  {Calarco}},\ }\href@noop {} {\bibfield  {journal} {\bibinfo  {journal}
  {Phys.\,Rev.\,A}\ }\textbf {\bibinfo {volume} {74}},\ \bibinfo {pages}
  {022712} (\bibinfo {year} {2006})}\BibitemShut {NoStop}%
\bibitem [{\citenamefont {Kestner}\ and\ \citenamefont
  {Sinano\v{g}lu}(1962)}]{qd:kest62}%
  \BibitemOpen
  \bibfield  {author} {\bibinfo {author} {\bibfnamefont {N.~R.}\ \bibnamefont
  {Kestner}}\ and\ \bibinfo {author} {\bibfnamefont {O.}~\bibnamefont
  {Sinano\v{g}lu}},\ }\href@noop {} {\bibfield  {journal} {\bibinfo  {journal}
  {Phys.\,Rev.}\ }\textbf {\bibinfo {volume} {128}},\ \bibinfo {pages} {2687}
  (\bibinfo {year} {1962})}\BibitemShut {NoStop}%
\bibitem [{\citenamefont {Stark}()}]{qd:star11}%
  \BibitemOpen
  \bibfield  {author} {\bibinfo {author} {\bibfnamefont {A.}~\bibnamefont
  {Stark}},\ }\href@noop {} {}\bibinfo {note} {Master thesis,
  Humboldt-Universt\"at zu Berlin, 2011}\BibitemShut {NoStop}%
\bibitem [{\citenamefont {Grishkevich}\ and\ \citenamefont
  {Saenz}(2009)}]{cold:gris09}%
  \BibitemOpen
  \bibfield  {author} {\bibinfo {author} {\bibfnamefont {S.}~\bibnamefont
  {Grishkevich}}\ and\ \bibinfo {author} {\bibfnamefont {A.}~\bibnamefont
  {Saenz}},\ }\href@noop {} {\bibfield  {journal} {\bibinfo  {journal}
  {Phys.\,Rev.\,A}\ }\textbf {\bibinfo {volume} {80}},\ \bibinfo {pages}
  {013403} (\bibinfo {year} {2009})}\BibitemShut {NoStop}%
\bibitem [{\citenamefont {Grishkevich}\ \emph {et~al.}(2011)\citenamefont
  {Grishkevich}, \citenamefont {Sala},\ and\ \citenamefont
  {Saenz}}]{cold:gris11}%
  \BibitemOpen
  \bibfield  {author} {\bibinfo {author} {\bibfnamefont {S.}~\bibnamefont
  {Grishkevich}}, \bibinfo {author} {\bibfnamefont {S.}~\bibnamefont {Sala}}, \
  and\ \bibinfo {author} {\bibfnamefont {A.}~\bibnamefont {Saenz}},\
  }\href@noop {} {\bibfield  {journal} {\bibinfo  {journal} {Phys.\,Rev.\,A}\
  }\textbf {\bibinfo {volume} {84}},\ \bibinfo {pages} {062710} (\bibinfo
  {year} {2011})}\BibitemShut {NoStop}%
\bibitem [{\citenamefont {Czycholl}(2008)}]{qd:czyc09}%
  \BibitemOpen
  \bibfield  {author} {\bibinfo {author} {\bibfnamefont {G.}~\bibnamefont
  {Czycholl}},\ }\enquote {\bibinfo {title} {{Theoretische
  Festk\"orperphysik}},}\ \ (\bibinfo  {publisher} {Springer, Berlin
  Heidelberg},\ \bibinfo {year} {2008})\ p.\ \bibinfo {pages} {196},\ \bibinfo
  {edition} {3rd}\ ed.\BibitemShut {Stop}%
\bibitem [{\citenamefont {Musial}\ \emph {et~al.}(2012)\citenamefont
  {Musial}, \citenamefont {Kaczmarkiewicz}, \citenamefont {S\c{e}k},
  \citenamefont {Podemski}, \citenamefont {Machnikowski},\ and\ \citenamefont
  {Misiewicz}}]{qd:musi12}%
  \BibitemOpen
  \bibfield  {author} {\bibinfo {author} {\bibfnamefont {A.}~\bibnamefont
  {Musial}}, \bibinfo {author} {\bibfnamefont {P.}~\bibnamefont
  {Kaczmarkiewicz}}, \bibinfo {author} {\bibfnamefont {G.}~\bibnamefont
  {S\c{e}k}}, \bibinfo {author} {\bibfnamefont {P.}~\bibnamefont {Podemski}},
  \bibinfo {author} {\bibfnamefont {P.}~\bibnamefont {Machnikowski}}, \ and\
  \bibinfo {author} {\bibfnamefont {J.}~\bibnamefont {Misiewicz}},\ }\href@noop
  {} {\bibfield  {journal} {\bibinfo  {journal} {Phys.\,Rev.\,B}\ }\textbf
  {\bibinfo {volume} {85}},\ \bibinfo {pages} {035314} (\bibinfo {year}
  {2012})}\BibitemShut {NoStop}%
\bibitem [{\citenamefont {Planelles}\ \emph {et~al.}(2009)\citenamefont
  {Planelles}, \citenamefont {Royo}, \citenamefont {Ballester},\ and\
  \citenamefont {Pi}}]{qd:plan09}%
  \BibitemOpen
  \bibfield  {author} {\bibinfo {author} {\bibfnamefont {J.}~\bibnamefont
  {Planelles}}, \bibinfo {author} {\bibfnamefont {M.}~\bibnamefont {Royo}},
  \bibinfo {author} {\bibfnamefont {A.}~\bibnamefont {Ballester}}, \ and\
  \bibinfo {author} {\bibfnamefont {M.}~\bibnamefont {Pi}},\ }\href@noop {}
  {\bibfield  {journal} {\bibinfo  {journal} {Phys.\,Rev.\,B}\ }\textbf
  {\bibinfo {volume} {80}},\ \bibinfo {pages} {045324} (\bibinfo {year}
  {2009})}\BibitemShut {NoStop}%
\bibitem [{\citenamefont {Tsukamoto}\ \emph {et~al.}(1992)\citenamefont
  {Tsukamoto}, \citenamefont {Nagamune}, \citenamefont {Nishioka},\ and\
  \citenamefont {Arakawa}}]{qd:tsuk92}%
  \BibitemOpen
  \bibfield  {author} {\bibinfo {author} {\bibfnamefont {S.}~\bibnamefont
  {Tsukamoto}}, \bibinfo {author} {\bibfnamefont {Y.}~\bibnamefont {Nagamune}},
  \bibinfo {author} {\bibfnamefont {M.}~\bibnamefont {Nishioka}}, \ and\
  \bibinfo {author} {\bibfnamefont {Y.}~\bibnamefont {Arakawa}},\ }\href@noop
  {} {\bibfield  {journal} {\bibinfo  {journal} {J.\,Appl.\,Phys.}\ }\textbf
  {\bibinfo {volume} {71}},\ \bibinfo {pages} {533} (\bibinfo {year}
  {1992})}\BibitemShut {NoStop}%
\bibitem [{\citenamefont {Chuang}(2009)}]{qd:chua09}%
  \BibitemOpen
  \bibfield  {author} {\bibinfo {author} {\bibfnamefont {S.~L.}\ \bibnamefont
  {Chuang}},\ }\enquote {\bibinfo {title} {{Physics of Photonic Devices}},}\ \
  (\bibinfo  {publisher} {John Wiley \& Sons, Inc. New Jersey},\ \bibinfo
  {year} {2009})\ pp.\ \bibinfo {pages} {802--803},\ \bibinfo {edition} {2nd}\
  ed.\BibitemShut {Stop}%
\bibitem [{\citenamefont {Zener}(1932)}]{cold:zene32}%
  \BibitemOpen
  \bibfield  {author} {\bibinfo {author} {\bibfnamefont {C.}~\bibnamefont
  {Zener}},\ }\href@noop {} {\bibfield  {journal} {\bibinfo  {journal}
  {Proc.\,R.\,Soc.\,A}\ }\textbf {\bibinfo {volume} {137}},\ \bibinfo {pages}
  {696} (\bibinfo {year} {1932})}\BibitemShut {NoStop}%
\bibitem [{\citenamefont {Landau}(1932)}]{cold:land32}%
  \BibitemOpen
  \bibfield  {author} {\bibinfo {author} {\bibfnamefont {L.}~\bibnamefont
  {Landau}},\ }\href@noop {} {\bibfield  {journal} {\bibinfo  {journal}
  {Phys.\,Z.\,Sowjetunion}\ }\textbf {\bibinfo {volume} {2}},\ \bibinfo {pages}
  {46} (\bibinfo {year} {1932})}\BibitemShut {NoStop}%
\bibitem [{\citenamefont {Wittig}(2005)}]{cold:witt05}%
  \BibitemOpen
  \bibfield  {author} {\bibinfo {author} {\bibfnamefont {C.}~\bibnamefont
  {Wittig}},\ }\href@noop {} {\bibfield  {journal} {\bibinfo  {journal}
  {J.\,Phys.\,Chem.\,B}\ }\textbf {\bibinfo {volume} {109}},\ \bibinfo {pages}
  {8428} (\bibinfo {year} {2005})}\BibitemShut {NoStop}%
\bibitem [{\citenamefont {'t~Hooft}\ \emph {et~al.}(1987)\citenamefont
  {'t~Hooft}, \citenamefont {van~der Poel}, \citenamefont {Molenkamp},\ and\
  \citenamefont {Foxon}}]{qd:hoof97}%
  \BibitemOpen
  \bibfield  {author} {\bibinfo {author} {\bibfnamefont {G.~W.}\ \bibnamefont
  {'t~Hooft}}, \bibinfo {author} {\bibfnamefont {W.~A. J.~A.}\ \bibnamefont
  {van~der Poel}}, \bibinfo {author} {\bibfnamefont {L.~W.}\ \bibnamefont
  {Molenkamp}}, \ and\ \bibinfo {author} {\bibfnamefont {C.~T.}\ \bibnamefont
  {Foxon}},\ }\href@noop {} {\bibfield  {journal} {\bibinfo  {journal}
  {Phys.\,Rev.\,B}\ }\textbf {\bibinfo {volume} {35}},\ \bibinfo {pages}
  {8281(R)} (\bibinfo {year} {1987})}\BibitemShut {NoStop}%
\bibitem [{\citenamefont {Hwang}(1973)}]{qd:hwan73}%
  \BibitemOpen
  \bibfield  {author} {\bibinfo {author} {\bibfnamefont {C.~J.}\ \bibnamefont
  {Hwang}},\ }\href@noop {} {\bibfield  {journal} {\bibinfo  {journal}
  {Phys.\,Rev.\,B}\ }\textbf {\bibinfo {volume} {8}},\ \bibinfo {pages} {646}
  (\bibinfo {year} {1973})}\BibitemShut {NoStop}%
\bibitem [{\citenamefont {Perera}\ \emph {et~al.}(2008)\citenamefont {Perera},
  \citenamefont {Fickenscher}, \citenamefont {Jackson}, \citenamefont {Smith},
  \citenamefont {Yarrison-Rice}, \citenamefont {Joyce}, \citenamefont {Gao},
  \citenamefont {Tan}, \citenamefont {Jagadish}, \citenamefont {Zhang},\ and\
  \citenamefont {Zou}}]{qd:pere08}%
  \BibitemOpen
  \bibfield  {author} {\bibinfo {author} {\bibfnamefont {S.}~\bibnamefont
  {Perera}}, \bibinfo {author} {\bibfnamefont {M.~A.}\ \bibnamefont
  {Fickenscher}}, \bibinfo {author} {\bibfnamefont {H.~E.}\ \bibnamefont
  {Jackson}}, \bibinfo {author} {\bibfnamefont {L.~M.}\ \bibnamefont {Smith}},
  \bibinfo {author} {\bibfnamefont {J.~M.}\ \bibnamefont {Yarrison-Rice}},
  \bibinfo {author} {\bibfnamefont {H.~J.}\ \bibnamefont {Joyce}}, \bibinfo
  {author} {\bibfnamefont {Q.}~\bibnamefont {Gao}}, \bibinfo {author}
  {\bibfnamefont {H.~H.}\ \bibnamefont {Tan}}, \bibinfo {author} {\bibfnamefont
  {C.}~\bibnamefont {Jagadish}}, \bibinfo {author} {\bibfnamefont
  {X.}~\bibnamefont {Zhang}}, \ and\ \bibinfo {author} {\bibfnamefont
  {J.}~\bibnamefont {Zou}},\ }\href@noop {} {\bibfield  {journal} {\bibinfo
  {journal} {Appl. Phys. Lett.}\ }\textbf {\bibinfo {volume} {93}},\ \bibinfo
  {pages} {053110} (\bibinfo {year} {2008})}\BibitemShut {NoStop}%
\bibitem [{\citenamefont {Champlin}\ and\ \citenamefont
  {Glover}(1968)}]{qd:cham68}%
  \BibitemOpen
  \bibfield  {author} {\bibinfo {author} {\bibfnamefont {K.~S.}\ \bibnamefont
  {Champlin}}\ and\ \bibinfo {author} {\bibfnamefont {G.~H.}\ \bibnamefont
  {Glover}},\ }\href@noop {} {\bibfield  {journal} {\bibinfo  {journal} {Appl.
  Phys. Lett.}\ }\textbf {\bibinfo {volume} {12}},\ \bibinfo {pages} {231}
  (\bibinfo {year} {1968})}\BibitemShut {NoStop}%
\bibitem [{\citenamefont {Bednarek}\ \emph {et~al.}(2003)\citenamefont
  {Bednarek}, \citenamefont {Szafran}, \citenamefont {Lis},\ and\ \citenamefont
  {Adamowski}}]{qd:bedn03}%
  \BibitemOpen
  \bibfield  {author} {\bibinfo {author} {\bibfnamefont {S.}~\bibnamefont
  {Bednarek}}, \bibinfo {author} {\bibfnamefont {B.}~\bibnamefont {Szafran}},
  \bibinfo {author} {\bibfnamefont {K.}~\bibnamefont {Lis}}, \ and\ \bibinfo
  {author} {\bibfnamefont {J.}~\bibnamefont {Adamowski}},\ }\href@noop {}
  {\bibfield  {journal} {\bibinfo  {journal} {Phys.\,Rev.\,B}\ }\textbf
  {\bibinfo {volume} {68}},\ \bibinfo {pages} {155333} (\bibinfo {year}
  {2003})}\BibitemShut {NoStop}%
\bibitem [{\citenamefont {Huber}\ \emph {et~al.}(1998)\citenamefont {Huber},
  \citenamefont {Zrenner}, \citenamefont {Wegschneider},\ and\ \citenamefont
  {Bichler}}]{qd:hube98}%
  \BibitemOpen
  \bibfield  {author} {\bibinfo {author} {\bibfnamefont {T.}~\bibnamefont
  {Huber}}, \bibinfo {author} {\bibfnamefont {A.}~\bibnamefont {Zrenner}},
  \bibinfo {author} {\bibfnamefont {W.}~\bibnamefont {Wegschneider}}, \ and\
  \bibinfo {author} {\bibfnamefont {M.}~\bibnamefont {Bichler}},\ }\href@noop
  {} {\bibfield  {journal} {\bibinfo  {journal} {Phys. Stat. Sol. A}\ }\textbf
  {\bibinfo {volume} {166}},\ \bibinfo {pages} {R5} (\bibinfo {year}
  {1998})}\BibitemShut {NoStop}%
\bibitem [{\citenamefont {Hammack}\ \emph
  {et~al.}(2006{\natexlab{a}})\citenamefont {Hammack}, \citenamefont {Gippius},
  \citenamefont {Yang}, \citenamefont {Andreev},\ and\ \citenamefont
  {Butov}}]{qd:hamm06}%
  \BibitemOpen
  \bibfield  {author} {\bibinfo {author} {\bibfnamefont {A.~T.}\ \bibnamefont
  {Hammack}}, \bibinfo {author} {\bibfnamefont {N.~A.}\ \bibnamefont
  {Gippius}}, \bibinfo {author} {\bibfnamefont {S.}~\bibnamefont {Yang}},
  \bibinfo {author} {\bibfnamefont {G.~O.}\ \bibnamefont {Andreev}}, \ and\
  \bibinfo {author} {\bibfnamefont {L.~V.}\ \bibnamefont {Butov}},\ }\href@noop
  {} {\bibfield  {journal} {\bibinfo  {journal} {J.\,Appl.\,Phys.}\ }\textbf
  {\bibinfo {volume} {99}},\ \bibinfo {pages} {0660104} (\bibinfo {year}
  {2006}{\natexlab{a}})}\BibitemShut {NoStop}%
\bibitem [{\citenamefont {High}\ \emph {et~al.}(2009)\citenamefont {High},
  \citenamefont {Thomas}, \citenamefont {Grosso}, \citenamefont {Remeika},
  \citenamefont {Hammack}, \citenamefont {Meyertholen}, \citenamefont {Fogler},
  \citenamefont {Butov}, \citenamefont {Hanson},\ and\ \citenamefont
  {Gossard}}]{qd:high09}%
  \BibitemOpen
  \bibfield  {author} {\bibinfo {author} {\bibfnamefont {A.~A.}\ \bibnamefont
  {High}}, \bibinfo {author} {\bibfnamefont {A.~K.}\ \bibnamefont {Thomas}},
  \bibinfo {author} {\bibfnamefont {G.}~\bibnamefont {Grosso}}, \bibinfo
  {author} {\bibfnamefont {M.}~\bibnamefont {Remeika}}, \bibinfo {author}
  {\bibfnamefont {A.~T.}\ \bibnamefont {Hammack}}, \bibinfo {author}
  {\bibfnamefont {A.~D.}\ \bibnamefont {Meyertholen}}, \bibinfo {author}
  {\bibfnamefont {M.~M.}\ \bibnamefont {Fogler}}, \bibinfo {author}
  {\bibfnamefont {L.~V.}\ \bibnamefont {Butov}}, \bibinfo {author}
  {\bibfnamefont {M.}~\bibnamefont {Hanson}}, \ and\ \bibinfo {author}
  {\bibfnamefont {A.~C.}\ \bibnamefont {Gossard}},\ }\href@noop {} {\bibfield
  {journal} {\bibinfo  {journal} {Phys.\,Rev.\,Lett.}\ }\textbf {\bibinfo
  {volume} {103}},\ \bibinfo {pages} {087403} (\bibinfo {year}
  {2009})}\BibitemShut {NoStop}%
\bibitem [{\citenamefont {Schinner}\ \emph {et~al.}(2013)\citenamefont
  {Schinner}, \citenamefont {Repp}, \citenamefont {Schubert}, \citenamefont
  {Rai}, \citenamefont {Reuter}, \citenamefont {Wieck}, \citenamefont
  {Govorov}, \citenamefont {Holleitner},\ and\ \citenamefont
  {Kotthaus}}]{qd:schi13}%
  \BibitemOpen
  \bibfield  {author} {\bibinfo {author} {\bibfnamefont {G.~J.}\ \bibnamefont
  {Schinner}}, \bibinfo {author} {\bibfnamefont {J.}~\bibnamefont {Repp}},
  \bibinfo {author} {\bibfnamefont {E.}~\bibnamefont {Schubert}}, \bibinfo
  {author} {\bibfnamefont {A.~K.}\ \bibnamefont {Rai}}, \bibinfo {author}
  {\bibfnamefont {D.}~\bibnamefont {Reuter}}, \bibinfo {author} {\bibfnamefont
  {A.~D.}\ \bibnamefont {Wieck}}, \bibinfo {author} {\bibfnamefont {A.~O.}\
  \bibnamefont {Govorov}}, \bibinfo {author} {\bibfnamefont {A.~W.}\
  \bibnamefont {Holleitner}}, \ and\ \bibinfo {author} {\bibfnamefont {J.~P.}\
  \bibnamefont {Kotthaus}},\ }\href@noop {} {\bibfield  {journal} {\bibinfo
  {journal} {Phys.\,Rev.\,Lett.}\ }\textbf {\bibinfo {volume} {110}},\ \bibinfo
  {pages} {127403} (\bibinfo {year} {2013})}\BibitemShut {NoStop}%
\bibitem [{\citenamefont {Sauerwald}\ \emph {et~al.}(2005)\citenamefont
  {Sauerwald}, \citenamefont {K\"ummell}, \citenamefont {Bacher}, \citenamefont
  {Somers}, \citenamefont {Schwertberger}, \citenamefont {Reithmaier},\ and\
  \citenamefont {Forchel}}]{qd:saue05}%
  \BibitemOpen
  \bibfield  {author} {\bibinfo {author} {\bibfnamefont {A.}~\bibnamefont
  {Sauerwald}}, \bibinfo {author} {\bibfnamefont {T.}~\bibnamefont
  {K\"ummell}}, \bibinfo {author} {\bibfnamefont {G.}~\bibnamefont {Bacher}},
  \bibinfo {author} {\bibfnamefont {A.}~\bibnamefont {Somers}}, \bibinfo
  {author} {\bibfnamefont {R.}~\bibnamefont {Schwertberger}}, \bibinfo {author}
  {\bibfnamefont {J.~P.}\ \bibnamefont {Reithmaier}}, \ and\ \bibinfo {author}
  {\bibfnamefont {A.}~\bibnamefont {Forchel}},\ }\href@noop {} {\bibfield
  {journal} {\bibinfo  {journal} {Appl. Phys. Lett.}\ }\textbf {\bibinfo
  {volume} {86}},\ \bibinfo {pages} {253112} (\bibinfo {year}
  {2005})}\BibitemShut {NoStop}%
\bibitem [{\citenamefont {Hammack}\ \emph
  {et~al.}(2006{\natexlab{b}})\citenamefont {Hammack}, \citenamefont
  {Griswold}, \citenamefont {Butov}, \citenamefont {Smallwood}, \citenamefont
  {Ivanov},\ and\ \citenamefont {Gossard}}]{qd:hamm06b}%
  \BibitemOpen
  \bibfield  {author} {\bibinfo {author} {\bibfnamefont {A.~T.}\ \bibnamefont
  {Hammack}}, \bibinfo {author} {\bibfnamefont {M.}~\bibnamefont {Griswold}},
  \bibinfo {author} {\bibfnamefont {L.~V.}\ \bibnamefont {Butov}}, \bibinfo
  {author} {\bibfnamefont {L.~E.}\ \bibnamefont {Smallwood}}, \bibinfo {author}
  {\bibfnamefont {A.~L.}\ \bibnamefont {Ivanov}}, \ and\ \bibinfo {author}
  {\bibfnamefont {A.~C.}\ \bibnamefont {Gossard}},\ }\href@noop {} {\bibfield
  {journal} {\bibinfo  {journal} {Phys.\,Rev.\,Lett.}\ }\textbf {\bibinfo
  {volume} {96}},\ \bibinfo {pages} {227402} (\bibinfo {year}
  {2006}{\natexlab{b}})}\BibitemShut {NoStop}%
\bibitem [{\citenamefont {Schulz}\ \emph {et~al.}(2015)\citenamefont {Schulz},
  \citenamefont {Sala},\ and\ \citenamefont {Saenz}}]{cold:schu15}%
  \BibitemOpen
  \bibfield  {author} {\bibinfo {author} {\bibfnamefont {B.}~\bibnamefont
  {Schulz}}, \bibinfo {author} {\bibfnamefont {S.}~\bibnamefont {Sala}}, \ and\
  \bibinfo {author} {\bibfnamefont {A.}~\bibnamefont {Saenz}},\ }\href@noop {}
  {\bibfield  {journal} {\bibinfo  {journal} {New J. Phys.}\ }\textbf {\bibinfo
  {volume} {17}},\ \bibinfo {pages} {065002} (\bibinfo {year}
  {2015})}\BibitemShut {NoStop}%
\end{thebibliography}

%

\end{document}